%% file: lmcmonit.tex
\documentstyle[graphicx]{mn}

\newcommand{\lmcone}{\mbox{LMC~X-1}}
\newcommand{\lmcthree}{\mbox{LMC~X-3}}
\newcommand{\Msun}{${\mbox{M}_\odot}$}
\newcommand\aproxgt{\mathrel{%
      \rlap{\raise 0.511ex \hbox{$>$}}{\lower 0.511ex \hbox{$\sim$}}}}
\newcommand\aproxlt{\mathrel{%
      \rlap{\raise 0.511ex \hbox{$<$}}{\lower 0.511ex \hbox{$\sim$}}}}

\begin{document}

\title{Discovery of Recurring Soft to Hard State Transitions in  LMC~X-3}
\author[J. Wilms et al.]{J.~Wilms$^1$, M.A.~Nowak$^2$,
  K.~Pottschmidt$^1$, W.A.~Heindl$^3$, 
  J.B.~Dove$^{4,5}$, \newauthor M.C.~Begelman$^{2,6}$ \\
$^1$ Institut f\"ur Astronomie und Astrophysik -- Astronomie,
  Waldh\"auser Str. 64, D-72076 T\"ubingen,
  Germany, \\ 
\{wilms,katja\}@astro.uni-tuebingen.de \\
$^2$ JILA, University of Colorado, Boulder, CO
  80309-440, U.S.A., \{mnowak,mitch\}@rocinante.colorado.edu \\
$^3$ Center for Astronomy and Space Sciences, Code
  0424, University of California at San Diego, La Jolla, CA 92093,
  U.S.A., \\
biff@ucsd.edu \\
$^4$ Center for Astronomy and Space Astrophysics,
  University of Colorado, Boulder, CO 80309-389,
  U.S.A.,dove@casa.colorado.edu \\
$^5$ also, Dept.\ of Physics, Metropolitan State College of
  Denver, C.B. 69, P.O. Box 173362, Denver, CO 80217-3362, U.S.A. \\
$^6$ also, Dept.\ of Astrophysics and Planetary Sciences,
  University of Colorado, Boulder 80309, U.S.A. }
\date{Submitted 2000 January, 1st revised version 2000 May}

\maketitle

\label{firstpage}

\begin{abstract}  
  We present the analysis of an approximately 3 year long Rossi X-ray
  Timing Explorer (RXTE) monitoring campaign of the canonical soft state
  black hole candidates \lmcone\ and \lmcthree. In agreement with previous
  observations, we find that the spectra of both sources can be
  well-described by the sum of a multi-temperature disk blackbody and a
  power law. In contrast to \lmcone, which does not exhibit any periodic
  spectral changes, we find that \lmcthree\ exhibits strong spectral
  variability on time scales of days to weeks. The variability pattern
  observed with the RXTE All Sky Monitor reveals that the variability is
  more complicated than the 99\,d or 198\,d periodicity discussed by Cowley
  et al. \shortcite{cowley:91a}.  For typical ASM count rates, the
  luminosity variations of \lmcthree\ are due to changes of the
  phenomenological disk blackbody temperature, $kT_{\rm in}$, between $\sim
  1$\,keV to $\sim 1.2$\,keV. During episodes of especially low luminosity
  (ASM count rates $\aproxlt 0.6\,\rm counts\,sec^{-1}$; four such periods
  are discussed here), $kT_{\rm in}$ strongly decreases until the disk
  component is undetectable, and the power law significantly hardens to a
  photon index of $\Gamma\sim 1.8$.  These changes are consistent with
  state changes of \lmcthree\ from the soft state to the canonical hard
  state of galactic black holes. We argue that the long term variability of
  \lmcthree\ might be due to a wind-driven limit cycle, such as that
  discussed by Shields et al. \shortcite{shields:86a}.
\end{abstract}

\begin{keywords}
accretion -- black hole physics -- Stars: binaries -- X-rays:
Stars -- Stars: LMC~X-1, LMC~X-3
\end{keywords}

\section{Introduction}\label{sect:intro}

Long term variability on time scales of months to years is seen in many
galactic black hole candidates. In analogy to the 35\,d cycle of
\mbox{Her~X-1}, the long term variability of some objects has been
identified with the precession of a warped accretion disk. Possible driving
mechanisms for a warp include radiation pressure due to the luminous
central X-ray source
\cite{pringle:96a,maloney:96a,maloney:97a,maloney:98a}, torques exerted by
an accretion disk wind \cite{schandl:94a,schandl:96a}, or tidal forces
(Larwood \nocite{larwood:98a} 1998, and references therein).

In some sources, the long term changes in the X-ray luminosity have been
associated with state changes of the accretion disk.  At low luminosities,
these sources are usually observed in the hard state, in which the X-ray
spectrum is dominated by a hard power-law component with a photon index of
$\Gamma\sim 1.7$ and an exponential rollover at $\sim 150$\,keV. The hard
state spectrum is usually described in terms of thermal Comptonization
(Sunyaev \& Tr\"umper \nocite{sunyaev:79a} 1979; Dove et
al. \nocite{dove:97c} 1997; and references therein).
At higher luminosities, black hole candidates exhibit a soft spectrum that
can be characterized by a (multi-temperature) blackbody with a peak
temperature of $kT\sim 1$\,keV.  In addition, a power law with a photon
index of $\Gamma\sim 2.5$ or softer is present. Changes between the
spectral states are typical for black holes as is evidenced by the 1996
soft state of \mbox{Cygnus~X-1} \cite{cui:96a,cui:98c}, and the
frequent transitions seen in GX~339$-$4
\cite{makishima:86a,bouchet:93a,wilms:98c,corbel:00a}.  See Tanaka \& Lewin
\shortcite{tanaka:95a}, Nowak \shortcite{nowak:95a}, and references therein
for further details.

Prior to the observations discussed here, \lmcone\ and \lmcthree\ were
the only black hole candidates that had always been seen in the soft
state.  Thus, both were ideal candidates for a systematic study of the
properties of the soft state.  Although both sources were otherwise
thought to be spectrally very similar, \lmcone\ had not shown signs of
any periodic long term variability, while \lmcthree\ was known to be
variable on a $\sim$100\,d time scale \cite{cowley:91a,cowley:94a}.
We therefore initiated a two resp.\ three year long campaign to
observe \lmcone\ and \lmcthree\ with the Rossi X-ray Timing Explorer
(RXTE).  During the campaign we performed X-ray observations of
roughly 10\,ksec length at approximately three week intervals. The
spacing and exposure time of the observations were chosen such that
they would enable us to track spectral changes of the sources over any
long term variability pattern. The campaign was started in
1996~December with two long ($\sim 170\,\rm ksec$) observations of the
sources.  The analysis of these observations is presented in a
companion paper (Nowak et al., 2000, henceforth
paper~I).\nocite{nowak:99f} Preliminary results from the first year of
this campaign, using earlier versions of the response matrix and
background models, have been presented elsewhere
\cite{wilms:99a,wilms:99c}, this paper is devoted to a discussion of
the first two years of the campaign.

The remainder of this paper is structured as follows.  In
Section~\ref{sec:rxte} we present the details of our data analysis
procedure. Section~\ref{sec:lmcx3} is devoted to the study of \lmcthree.
Section~\ref{sec:lmcx1} contrasts these observations with those from
\lmcone. In Section~\ref{sec:softphys} we interpret our results in the
context of current models for the soft state of galactic black hole
candidates, and in the context for models of the long term variability of
X-ray binaries. We summarize our results in Section~\ref{sec:summary}.
Throughout this paper we assume a distance to the LMC of 50\,kpc.

\section{Data Analysis}\label{sec:rxte}

Onboard RXTE are two pointed instruments--- the Proportional Counter
Array (PCA) and the High Energy X-ray Timing Experiment (HEXTE)--- as
well as the All Sky Monitor (ASM). We used the standard RXTE data
analysis software, ftools~4.2, to examine the PCA and HEXTE data.
Spectral modeling was done using \textsc{xspec}, version 10.00ab
\cite{arnaud:96a}.  Due to the short duration of the pointed
observations we use only the PCA data for this analysis.  We used
essentially the same data screening and analysis strategy as in
paper~I, i.e., we used PCA top xenon layer data only and ignored data
taken within 30 minutes after passages through the south Atlantic
anomaly and where the background count rate, as measured by the
``electron ratio'', was comparatively large (see paper~I for details).
Data from all PCUs was combined for the final analysis.  Contrary to
paper~I, the observations presented here are so short that the Poisson
error dominates the uncertainty of the spectrum; therefore, no
systematic error was applied to the data.  For some of the
observations, only part of the PCA detectors were turned on. These
intervals were extracted separately and then combined. Response
matrices were generated for each of these intervals. The response
matrix for the final analysis was obtained from these individual
matrices by adding the matrices weighted by the fraction of photons
coming from each of the intervals.  We analyzed data taken in the
energy band from 2.5\,keV to 20\,keV.

Since our data are background dominated above $\sim$5\,keV, good background
modeling is essential for our analysis.  Background subtraction of the PCA
data was performed using a model taking into account sky pointings of the
PCA, and modeling the background variability using data from the Very Large
Event (VLE) counter of the PCA or using the so-called ``Faint model'',
depending on the source count rate.  To estimate the uncertainty of the
background model we compared the count rate measured in the PCA channels
above 30\,keV, where no source photons are detected, with the background
model flux at these energies. In all cases the agreement between the
background model and the measured background was within 2\%. We accounted
for this fluctuation by renormalizing the background model flux, using the
\textsc{xspec} \textit{corrfile} facility, such that $\chi^2$ was minimized
in our spectral fits.  Note that this introduces a ``hidden'' fitting
parameter to the data analysis.  This parameter was not taken into account
in the number of degrees of freedom since we assume that it converges to
the correct normalization of the background model, and therefore does not
introduce an uncertainty into the spectral fitting process.

We used data from the ASM in order to be able to place our observations in
the context of the long term variability of the sources.  This instrument,
an array of shadow cameras scanning the whole sky visible from the
spacecraft for five to ten times per day \cite{remillard:97a,levine:96a},
provides almost uninterrupted information about the long term behaviour of
bright X-ray sources.  We used the ``definitive one-dwell'' data available
from the Goddard Space Flight Center (GSFC) and only used flux solutions
for which $\chi_{\rm red}^2<1.1$.

\section{LMC X-3: Long Time Scale Spectral Variability}\label{sec:lmcx3}

\subsection{Introduction}
The black hole candidate \lmcthree\ was discovered during
\textsl{UHURU} observations of the Large Magellanic Cloud
\cite{leong:71a}.  A summary of the early observational history has
been given by Treves et al.  \shortcite{treves:88a}.  Interest in
\lmcthree\ was heightened when it was realized that its X-ray spectrum
is very similar to that of \mbox{Cyg~X-1} in the high state.  White \&
Marshall \shortcite{white:84a} therefore established the source as a
potential black hole candidate. Subsequent radial velocity
measurements, resulting in a mass function of 2.3\,\Msun
\cite{cowley:83a}, as well as optical photometry (van der Klis et al.
\nocite{vanderklis:83a} 1983; Kuiper et al. \nocite{kuiper:88a} 1988;
and references therein) of the B3V companion lead to a minimum mass of
$\sim$4\,\Msun \cite{mazeh:86a}, with a most probable mass of
9--10\,\Msun\ \cite{cowley:83a,paczynski:83a} for the compact object.
\lmcthree\ thus almost certainly contains a black hole.

The luminosity of \lmcthree\ can be as much as $\sim$30\% of the
Eddington luminosity of a 9\,\Msun\ black hole.  The source has been
known to be strongly variable since \textsl{Ariel V} and the High
Energy Astrophysics Observatory~1 (\textsl{HEAO~1}) discovered large
($>5$) intensity changes on time scales of weeks to months
\cite{griffiths:77a,johnston:79a}.  A first systematic study of its
long term periodicity was performed by Cowley et al.
\shortcite{cowley:91a}, who, based on \textsl{Ginga} and
\textsl{HEAO~1} data, found a long term period of 98.9\,d or 197.8\,d.
In addition, optical photometry indicated a lag of $\sim 20$\,d
between the optical and the X-rays (the optical leading the X-rays;
Cowley et al.  \nocite{cowley:91a} 1991).  Later Hubble Space
Telescope (HST) investigations revealed the presence of the
periodicity also in the ultraviolet \cite{cowley:94a}, albeit the long
term trends were not as clear as in the earlier data.  These HST
observations and the shape of the X-ray lightcurve led Cowley et al.
\shortcite{cowley:94a} to prefer the $\sim 99$\,d period and to
postulate that \lmcthree\ might have gone through a low-luminosity
state similar to the ``extended low'' states that are seen in Her~X-1
\cite{parmar:85a}.

Detailed spectral studies of the long term variability were first
performed by Ebisawa et al. \shortcite{ebisawa:93a}, who used data
from \textsl{Ginga} LAC pointings, taken over a period of three years,
that covered a large range of source luminosities.  Ebisawa et al.
\nocite{ebisawa:93a} showed that the X-ray spectra could be
well-described with a multi-temperature disk blackbody model plus a
power-law spectrum.  Alternatively, optically thick ($\tau_{\rm
  e}\aproxgt 20$) Comptonization could be used (see also Treves et al.
\nocite{treves:88a} 1988), but is not preferred due to the unphysical
parameters resulting from the spectral fitting. In accordance with
Cowley et al. \shortcite{cowley:91a}, Ebisawa et al.
\shortcite{ebisawa:93a} found that the soft spectral component hardens
with increasing source luminosity (i.e., its temperature inceases),
while the power-law tail was found to be independent of the
luminosity. One possible interpretation of this result is a varying
mass accretion rate $\dot{M}$.

A major shortcoming of the earlier spectral observations is that it
was impossible to put the observations in the context of the overall
long term variability of the source due to the lack of a real all-sky
monitoring instrument.  Furthermore, the relative inflexibility of the
early X-ray satellites made it impossible to obtain an even sampling
of X-ray spectra over the long time scale changes observed. A
satellite such as \textsl{RXTE}, which combines a broad spectral
coverage with very flexible scheduling and an all-sky monitoring
instrument, is therefore ideal to systematically study the long term
behaviour of sources such as \lmcthree.  In the remainder of this
section we discuss the \textsl{RXTE}-ASM soft X-ray lightcurve
(\S\ref{sec:x3asm}), describe our analysis of the pointed data
(\S\ref{sec:x3pca}), and present the results of our campaign
(\S\ref{sec:x3var}).

\begin{figure}
\begin{center}
\includegraphics[width=0.5\textwidth]{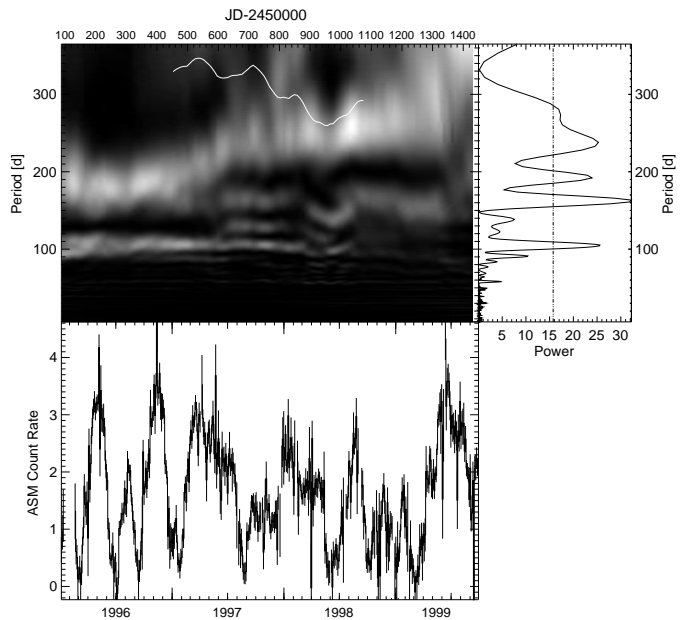}
\end{center}
\caption{{\bf a)} {\it Upper left:} Dynamical PSD of \lmcthree\ for
  the ASM data from 1996 through mid 1999. The white line displays the
  behaviour of the average ASM count-rate (a period of 350\,d
  corresponds to 4\,ASM cps).  {\bf b)} {\it Upper Right:}
  Lomb-Scargle-Periodogram for the ASM data through 1999, the dashed
  line indicates a false alarm probability for periods at the 0.01\%
  level (for the definition of the false alarm probability, see
  Scargle 1982). {\bf c)} {\it Lower left:} ASM lightcurve of
  \lmcthree, binned to a temporal resolution of 3\,d and variation of
  the spectral parameters for our RXTE pointed observations.  See text
  for explanations.  \protect{\label{fig:x3dynscarg}}}
\end{figure}

\subsection{Long Term Variability of LMC X-3}\label{sec:x3asm}

\begin{figure*}
\begin{center}
  \includegraphics[width=0.75\textwidth]{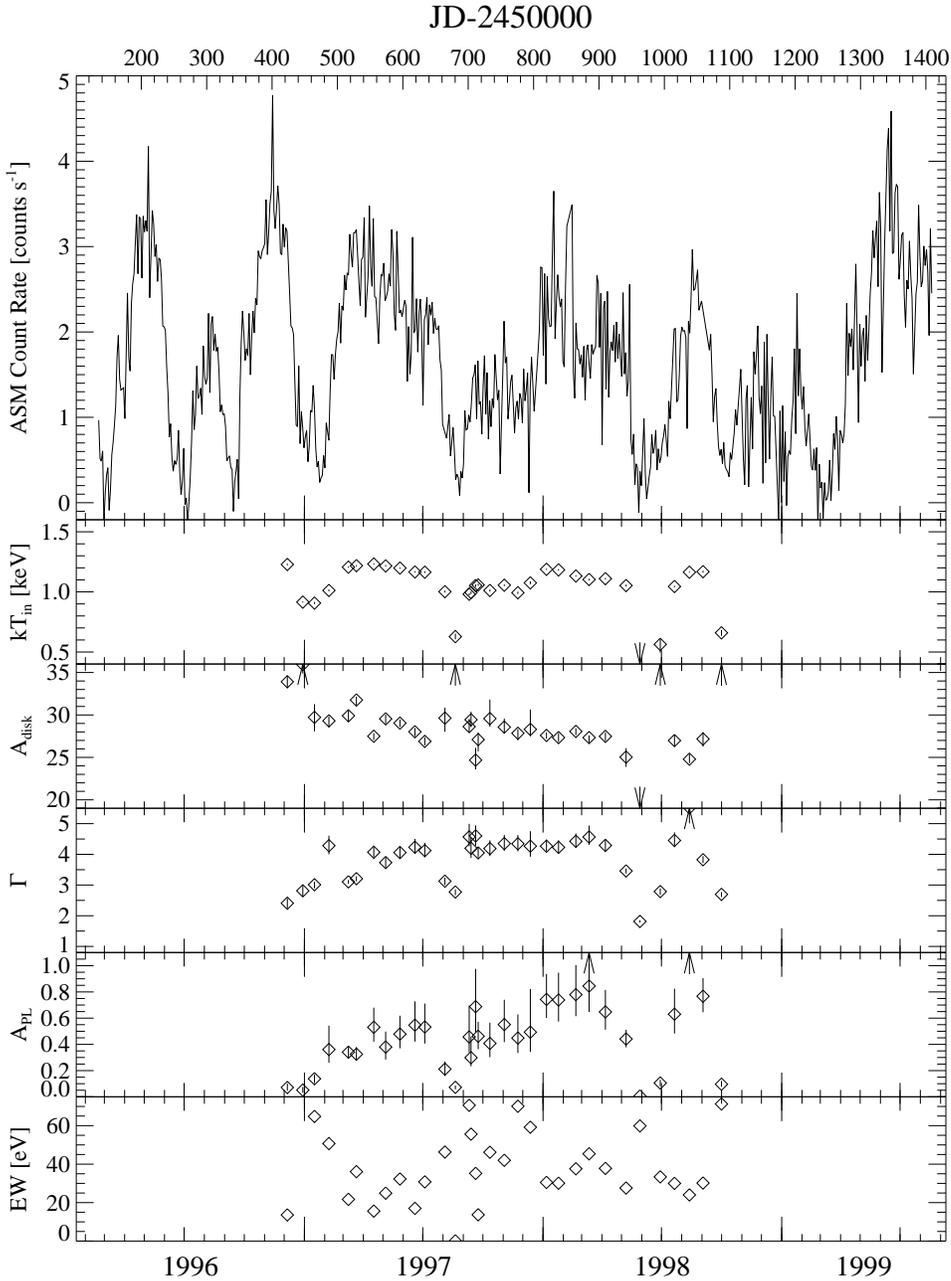}
\end{center}
\caption{Temporal variability of the spectral parameters of
  \lmcthree.}\label{fig:x3temp} 
\end{figure*}

In Fig.~\ref{fig:x3dynscarg}c we display the long term light curve of
the object as observed with the ASM, binned to a resolution of 3\,d.
Note the strong variability on time scales of $\sim 100$ and $\sim
250$\,d, which is well sampled by our monitoring (see
Fig.~\ref{fig:x3temp}). This variability had already been noted during
the first year of ASM operations \cite{levine:96a}.
 
The variability pattern itself is very complicated. The analysis of the ASM
light curve using the generalized periodogram of Lomb \shortcite{lomb:76a}
and Scargle \shortcite{scargle:82a} reveals significant periodicities on
long time scales of $\sim 100$, $\sim 160$, $\sim 190$, and $\sim
240$\,days (Fig.~\ref{fig:x3dynscarg}b). Epoch folding analysis
\cite{leahy:83a,schwarz:89a,davies:90a} independently verifies this result.
No significant variability (using the full temporal resolution ASM
lightcurve) is detected by these methods on the 1.7\,d orbital time scale
of the system. This is consistent with earlier results
\cite{weisskopf:83a,cowley:83a,paradijs:87a}, but contrary to other high
mass X-ray binaries, such as \mbox{Cyg~X-1} or \mbox{Vela~X-1}, where a
clear orbital modulation of the X-ray flux is detected in the ASM.  For
these latter sources, this variation is most probably due to photoelectric
absorption in the stellar wind \cite{wen:99a}.  The lack of orbital
variability in \lmcthree\ can therefore be interpreted as a lack of a
strong stellar wind from the companion star. This is not very surprising,
considering that the companion is a B3 V star \cite{warren:75a,cowley:94a}.

Compared to sources with clear perioditicies such as \mbox{Her~X-1}, the
peaks in the Lomb-Scargle periodogram of Fig.~\ref{fig:x3dynscarg}b are
quite broad. This can be a sign that the periods detected by this method
are only quasi-periodic, i.e., they change with time or are not present
during the whole time span covered. We therefore computed a dynamical
Lomb-Scargle periodogram by taking slices of the ASM light curve (binned
to a resolution of 3\,d) with a length of 730\,d each, and shifting these
slices with a step size of 9\,d over the whole ASM data range\footnote{In
  order to extend the dynamical periodogram to the full length of time
  covered by the ASM data, we added gaussianly distributed random data
  spanning one year before and after the ASM light curve. Each fake data
  set was chosen to have the same mean and variance of the first and last
  year, respectively, of the actual measured data set.}.  For each of these
slices the Lomb-Scargle periodogram was computed for the period range from
6\,d to 365\,d. To ensure that the periodograms are comparable, we
normalized them such that
\begin{equation}
\frac{\sigma^2}{\mu}=\int_{f_{\rm min}}^{f_{\rm max}}
{\rm PSD}(f)\,{\rm d}f ~~~,
\end{equation}
where $f_{\rm min}$ and $f_{\rm max}$ are the minimum and maximum frequency
for which the Lomb-Scargle periodogram is computed, and where $\sigma^2$ is
the variance and $\mu$ the mean ASM count rate of the 730\,d lightcurve.
The resulting periodogram was then gray scale coded and is displayed in
Fig.~\ref{fig:x3dynscarg}a, with each periodogram being displayed at the
mid-time of the light curve for which it was computed.

Although the individual Lomb-Scargle periodograms are obviously not
statistically independent, our approach is useful in revealing the
long term trends of the light curve. Fig.~\ref{fig:x3dynscarg}a shows
clearly the origin of the individual peaks in the total periodogram of
the source.  Albeit with varying significance, the dynamical
periodogram shows that a $\sim 100$\,d periodicity is present in all
data segments analyzed through the middle of 1998, confirming the
periodicity discussed by Cowley et al.  \shortcite{cowley:94a}.  The
$\sim 100$\,d periodicity is mainly attributable to the times of low
($<0.5$\,ASM cps) source luminosity, that are, e.g., seen in 1996
July, 1997 January, and 1997 August (Fig.~\ref{fig:x3dynscarg}c).
After mid 1998, the amplitude of the 100\,d periodicity is seen to
decrease; however, some of this reduction in amplitude can be
attributed to the increasing inclusion of fake data in a given two
year periodogram.  A possible shift in the distance between the light
curve minima during mid 1998 through 1999 may lead to the additional
time scale of $\sim 160$\,d present in Fig.~\ref{fig:x3dynscarg}b
(although see the discussion below). The ASM lightcurve for 1996 looks
similar to the folded \textsl{HEAO~1} and \textsl{Ginga} lightcurve
presented by Cowley et al.  \shortcite{cowley:91a}.  This might be an
indication that the ``quasi-sinusoidial'' variability of 1996 is not
just a random event, but occurs frequently.

The peaks at lower frequency in the total Lomb-Scargle periodogram are
due to long term periodicities that are not fixed in period. The
dynamical Lomb-Scargle periodogram shows a dominant periodicity on a
time scale of $\sim 190$\,d in 1996, which increases to $\sim 270$\,d
for the 730\,d light curve centered on 1997 December.  These shifts in
period can be attributed to the maxima of the ASM light curve, which
behave quasi-sinusoidally in 1996 and then show an increase in period
in 1997 and the first half of 1998, before (possibly) returning to a
more periodic behaviour\footnote{Note that the determination of
  periods $\aproxgt 250$\,d in the dynamical Lomb-Scargle periodogram
  is difficult since it is based on segments of 730\,d length only.
  Since the long period shows up in the total periodogram of
  Fig.~\ref{fig:x3dynscarg}b, however, we believe the dynamical
  periodogram to represent at least the general trends in the period
  shifting behaviour.}.  The strength of the individual peaks in the
total periodogram (Fig.~\ref{fig:x3dynscarg}b) can thus be seen as
partly depending on the duration of the intervals in which each period
dominated.  It is also possible that the $\sim 160$\,d peak seen in the
periodogram is the continuation of the $\sim 190$\,d period seen at
the start of 1996, while the $\sim 240$\,d peak is due to the
formation of a new periodicity beginning in late 1996 or early 1997.
It is also interesting to note that the $\sim 100$\,d and $\sim
190$\,d periodicities that exist at the beginning of the ASM light
curve have approximately the same period separation as the $\sim 160$
and $\sim 240$\,day periods that exist at the end of the ASM light
curve.

Although the ASM does in principle provide X-ray colour information,
\lmcthree\ is too weak to enable us to detect any statistically significant
changes in the spectral shape during the three years of ASM coverage.
Usable X-ray colours can only be computed after rebinning the data on
$\aproxgt 4$\,d intervals. These data indicate that the source might harden
during lower ASM count rates.  With the ASM, however, it is not possible to
characterize these changes in greater detail. We therefore turn to our
pointed monitoring observations to further describe these spectral trends.

\input{table1.tex}

\subsection{Spectral Modeling of \lmcthree\ PCA Data}\label{sec:x3pca}

Using the screening criteria of section~\ref{sec:rxte}, we extracted the
data from our 1997 and 1998 observations. A log of the observations is
given in Table~\ref{tab:x3log}.  Our conservative screening criteria reduce
the effective usable exposure times from the scheduled 10\,ksec to the
times shown in the Table.

\begin{figure}
\begin{center}
\includegraphics[width=0.45\textwidth]{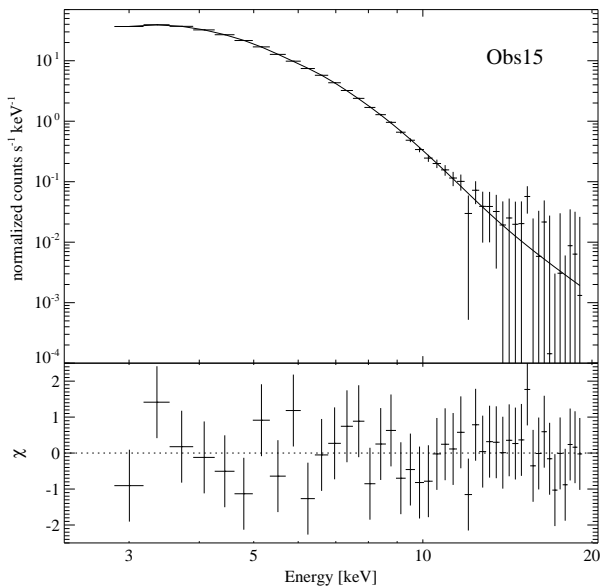}
\end{center}
\caption{Typical fit of a multi-temperature disk spectrum and a
  power law to data from \lmcthree. The bottom panel shows the
  residuals in units of $\sigma$. The residuals are consistent
  with the remaining calibration uncertainty of the PCA\label{fig:x3typ}.}
\end{figure}

We modeled the RXTE spectra using the standard multi-temperature disk
model \cite{mitsuda:84a,makishima:86a} plus an additional power law
component.  This is the traditional spectral model for describing the
X-ray continuum of the soft state.  We further discuss below the
interpretation and possible shortcomings of this model
(\S\ref{sec:softphys}, see also paper~I).  Our fits generally gave
acceptable results. Several observations, however, showed evidence for
deviations of the residuals in the iron K$\alpha$ band. Although these
deviations are mainly in the 2$\sigma$ range, adding a narrow
($\sigma=0.1$\,keV) iron line to the model improved our fits. We
therefore added a narrow Gaussian line feature at 6.4\,keV to all of
our fits.  Although the presence of a line is consistent with our
analysis from paper~I, we note that Fe line studies with the PCA are
complicated by the presence of systematic features in the PCA response
matrix \cite{wilms:98c} and by the fact that the power-law and the
disk model have comparable flux in the Fe-line region.  The line
parameters that we found here, therefore, should be seen as consistent
with an upper limit of $\sim$60\,eV for the equivalent width.  We took
the absorption in the intervening interstellar medium into account by
fixing the equivalent hydrogen column to the value found from radio
and ASCA observations $N_{\rm H}=3.2\times 10^{20}\,\rm cm^{-2}$
(Lister-Staveley, 1999, priv.\ comm.; see also paper~I), using the
cross sections of Ba\l{}uci\'{n}ska-Church \& McCammon
\shortcite{balu:92a}.  Using a fixed $N_{\rm H}$ is justified since
preliminary modeling showed that none of the observations had the
large $N_{\rm H}$ values that are detectable with the PCA energy range
and spectral resolution ($N_{\rm H}\aproxgt 10^{22}\,\rm cm^{-2}$;
Stelzer et al.  \nocite{stelzer:98a} 1999).

We list the results of our spectral modeling in Tab.~\ref{tab:x3fits}, and
a typical spectrum is shown in Fig.~\ref{fig:x3typ}.  Generally, the
spectral model described the data satisfactorily with reduced $\chi^2$
values of $\chi^2_{\rm red} \aproxlt 2$. We note that especially for some
brighter observations, the formal $\chi^2$ values are influenced by
systematic features in the response matrix. This can be examined, e.g., by
comparing the ratio between the data and the best fit model, with the ratio
between an observation of the Crab pulsar and nebula and the best fit power
law to that observation \cite{wilms:98c}. Visual inspection reveals that
the deviations between the data and the model for all monitoring
observations were comparable to or smaller than those seen in the Crab.
Thus, although $\chi^2_{\rm red}>1$ in these cases, adding additional
spectral components to the model would only serve to model the response
matrix features and not the observed spectrum.  We are therefore confident
that, given the spectral resolution of the PCA and the faint flux level of
\lmcthree, we have found a satisfactory spectral description for all
pointed observations.

We emphasize that the interpretation of the spectral components in
terms of physical parameters, however, is hampered by the properties
of the multi-temperature disk blackbody model. This discussion is
necessitated by the result that the typical power law indices found in
our data analysis are softer than those found by \textsl{Ginga}
($\Gamma \sim 2.2$; Ebisawa et al.  \nocite{ebisawa:93a} 1993). A more
detailed discussion can be found in paper~I.  A problem associated
with using the multi-temperature disk blackbody plus a power law for
modeling the soft state is that especially for large $\Gamma$ (i.e.,
soft power-law tail), the power law can have a non-negligible flux in
the lowest pulse height analyzer (PHA) channels so that the fit
parameters of the power law are almost completely determined by these
low channels.  This effect is present in most of our fits for which
$\Gamma\aproxgt 4$.  We tried to limit the influence of the power law
in the higher PHA channels with several other spectral models, e.g.,
by adding a heavily absorbed power-law to the disk blackbody, by using
a modified multi-temperature disk blackbody which includes a high
energy power law tail only above a given energy, or by applying
thermal Comptonization models.  None of these models resulted in
acceptable $\chi^2$ values.  In addition, using the multi-temperature
disk blackbody and leaving the equivalent column $N_{\rm H}$ a free
fit parameter always resulted in $N_{\rm H}\rightarrow 0\,\rm cm^2$.

The only model, apart from the one that we adopted here, which allowed
us to describe the spectrum of \lmcthree\ with similar or even better
$\chi^2$ values was a multi-temperature disk blackbody to which a low
energy blackbody spectrum was added. Such a model has been successfully
applied, e.g., also to the 1997 RXTE observations of GRO J1655$-$40
\cite{rothschild:99a}. Although such a model works for those observations
of \lmcthree\ where we find $\Gamma\aproxgt 4$, the model fails to reproduce
those observations where a harder power law component is clearly present in
the data. We therefore decided to use the multi-temperature disk blackbody
spectrum plus a power law in our analysis.

The significant contribution of the power law in the low-energy PHA
channels has some effect on the determination of the characteristic
disk temperature, $kT_{\rm in}$. As we noted in paper~I, the blackbody
temperature obtained from ASCA data was systematically lower than the
$kT_{\rm in}$ derived from the PCA data.  We speculated that this was
due to the significance of the power law in the low energy PHA
channels and the differing low energy cutoffs between the two
detectors.  We further speculated that the ASCA fit temperature was
more characteristic of the temperature for the ``seed photons'' that
are Compton upscattered to become the high energy spectrum.  As we
further discuss below for our observations of \lmcone\ 
(\S\ref{x1:longterm}), the presence of the power law may greatly
affect the determination of the blackbody normalization, $A_{\rm
  disk}$.

\input{table2.tex}

\subsection{Transitions between the Soft and Hard State}\label{sec:x3var}

\begin{figure*}
\begin{center}
\includegraphics[width=0.4\textwidth]{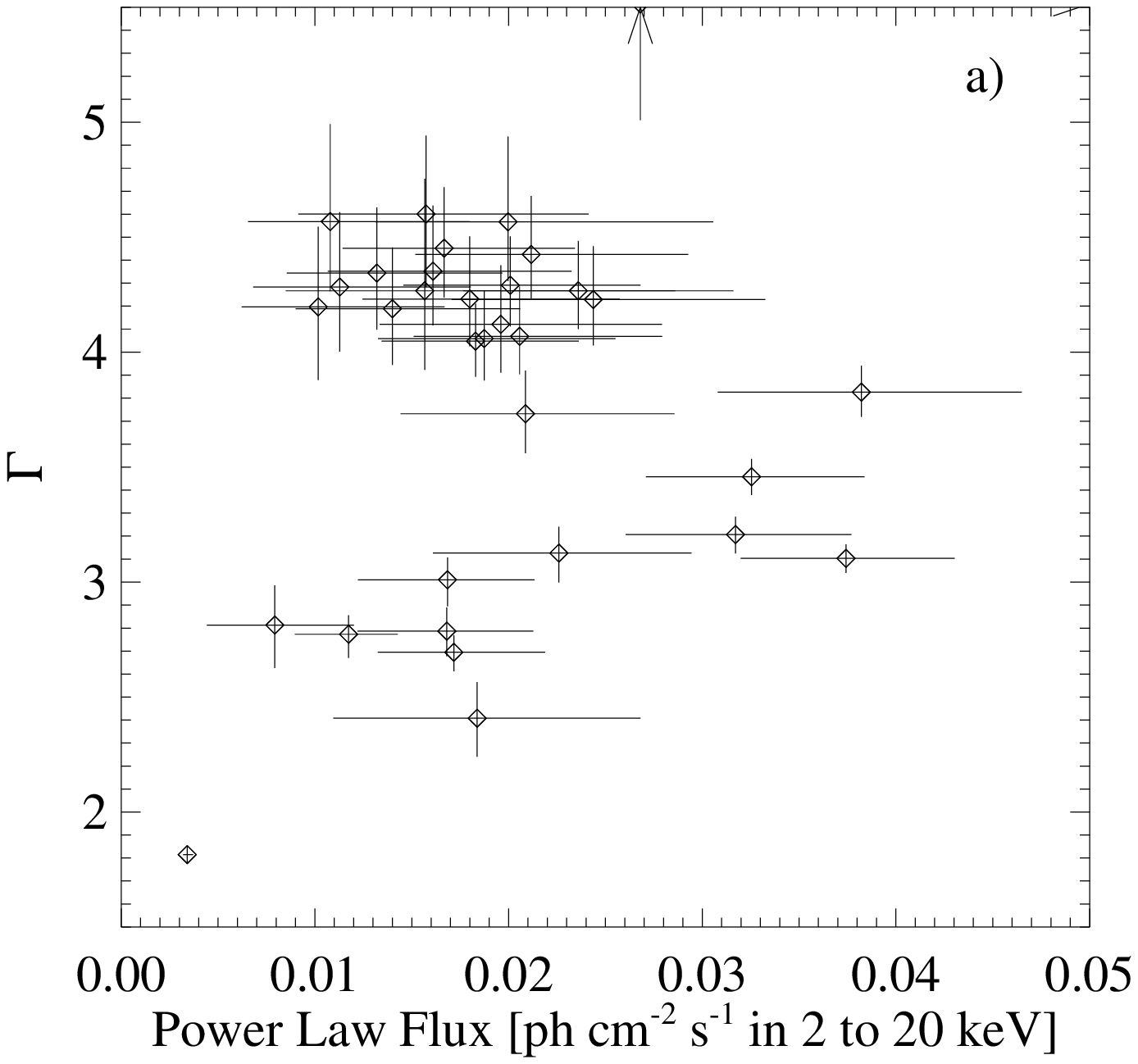}
\includegraphics[width=0.4\textwidth]{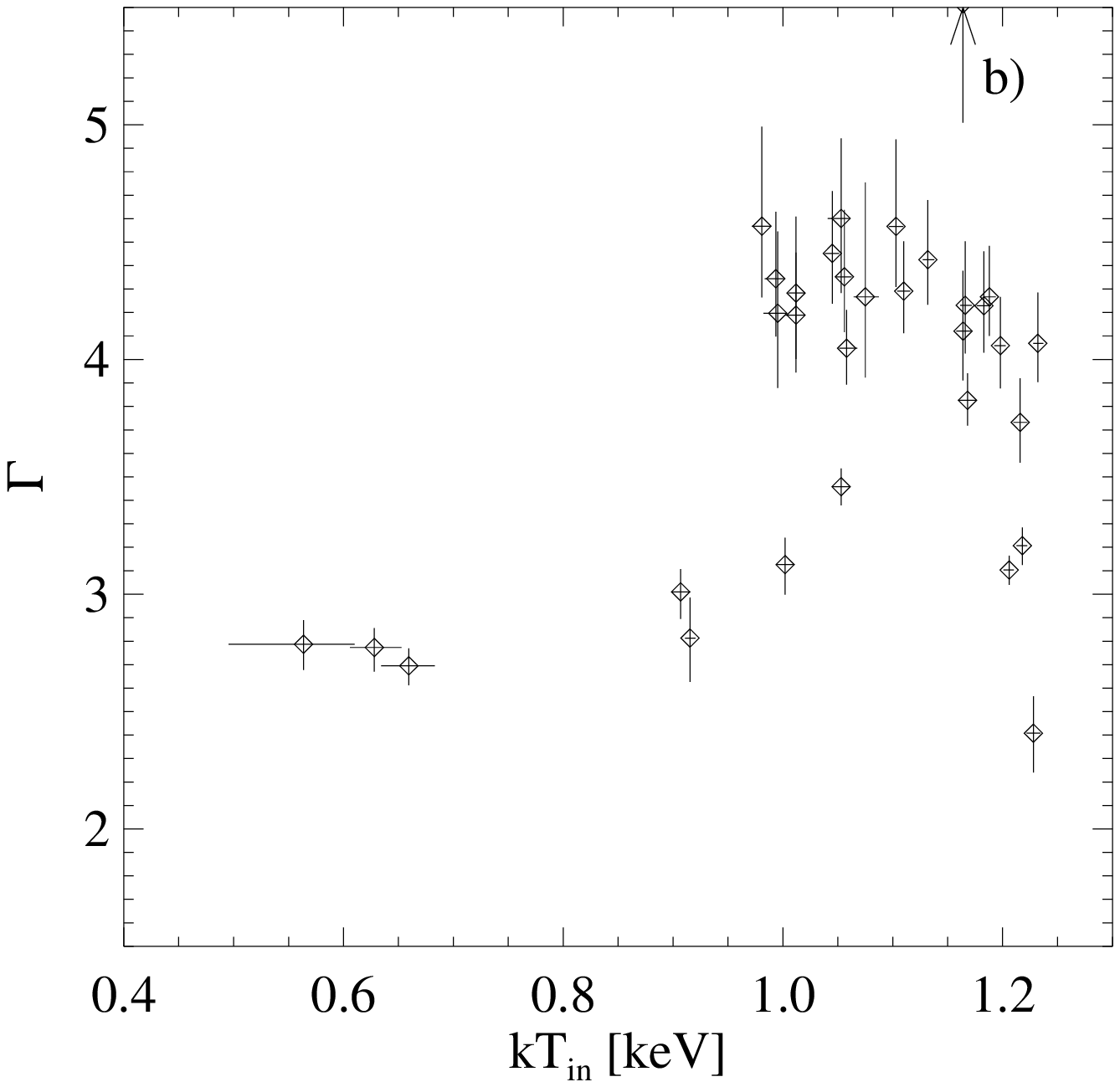}

\includegraphics[width=0.4\textwidth]{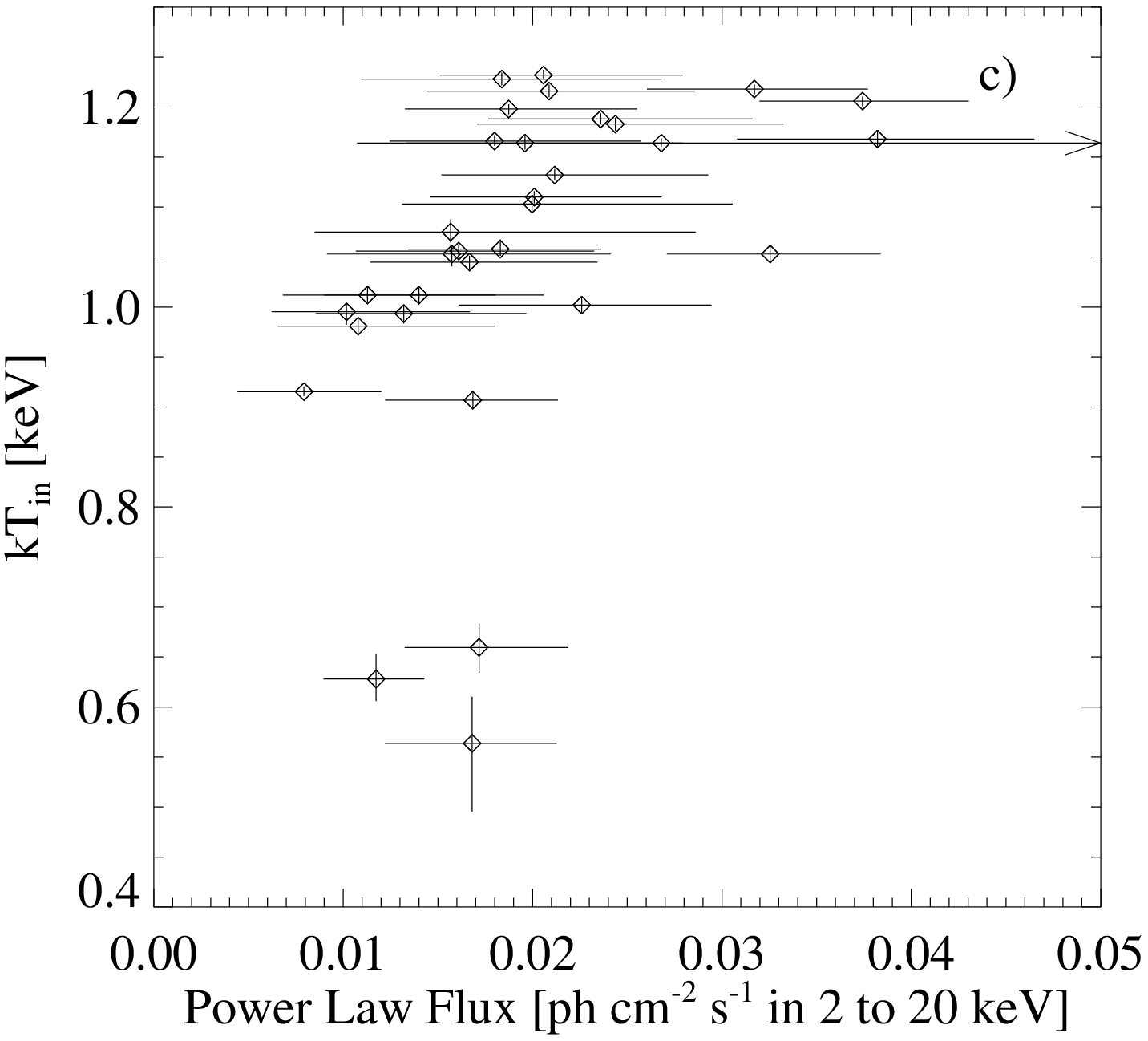}
\includegraphics[width=0.4\textwidth]{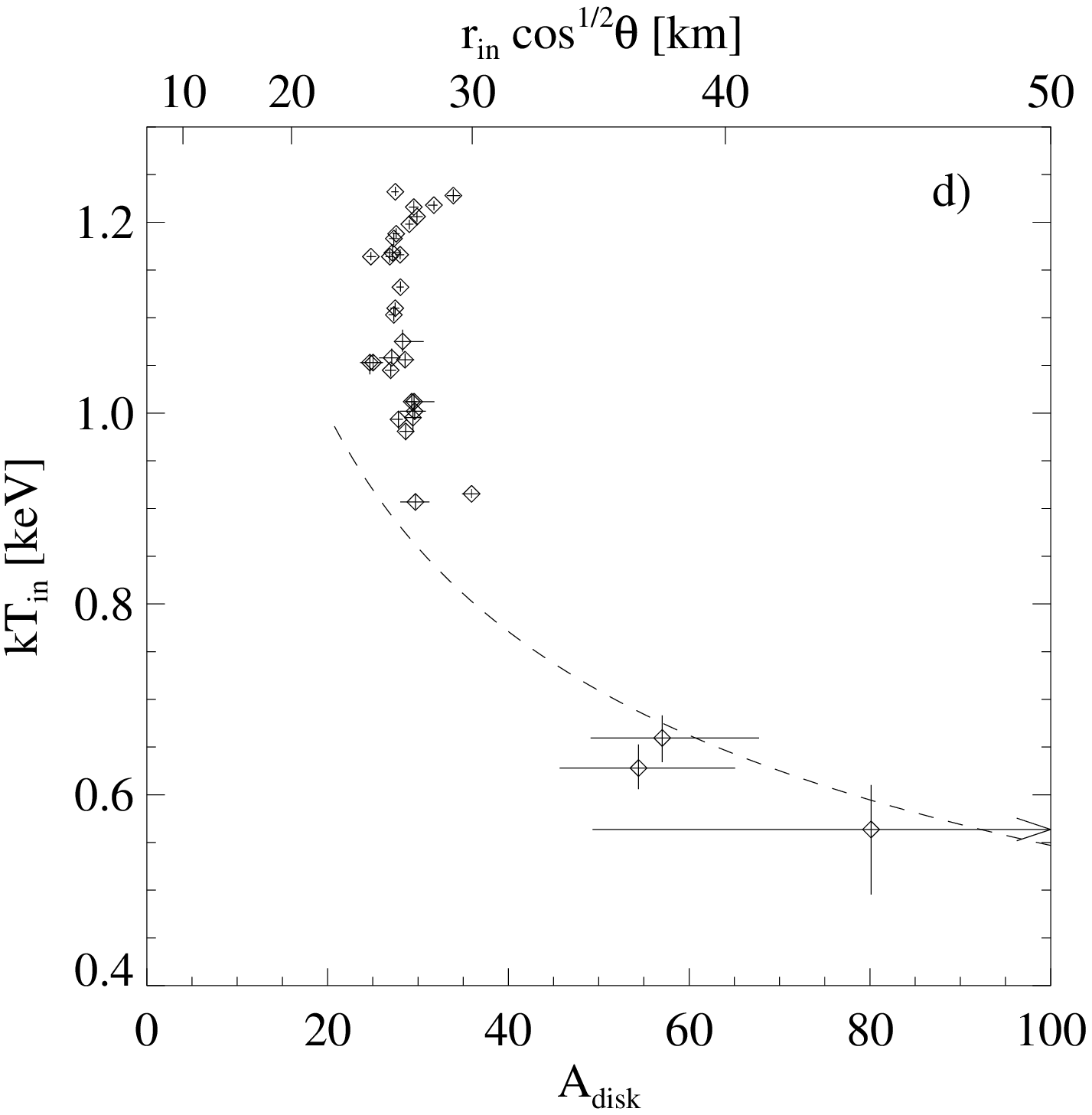}
\end{center}
\caption{Correlation between the spectral parameters for
  \lmcthree. The dashed line in subfigure d denotes the
  proportionality $A_{\rm disk} \propto kT_{\rm in}^{-8/3}$ expected
  for $T_{\rm in} \propto r_{\rm in}^{-3/4}$ (see text).}\label{fig:x3cor}
\end{figure*}

We now discuss the variations of the spectral parameters in terms of the
long term variability of \lmcthree.  In Fig.~\ref{fig:x3temp} we display
the spectral variation of \lmcthree\ in the context of its ASM long term
light curve.  The plot reveals a clear correlation between the total source
luminosity, as expressed by the ASM count rate, and the parameters of the
soft spectral component. During times of higher source luminosity, the
temperature of the multi-temperature disk blackbody, $kT_{\rm in}$,
increases. At the same time, the normalization of the disk blackbody stays
remarkably constant (see also Fig.~\ref{fig:x3cor}d). This effect is
similar to that found in the soft-state of other galactic black holes
\cite{tanaka:95a}. During episodes of comparably low ASM
count rate, however, the spectrum shows clear deviations from the typical
pure soft state behaviour. During our campaign, there were four of these
events: in 1997 January, 1997 August, 1998 June, and 1998 October. During
these times $kT_{\rm in}$ decreased to $kT_{\rm in}\aproxlt 0.5$\,keV,
while the photon index hardened (Fig.~\ref{fig:x3cor}b). For the most
extreme case, Obs28, no evidence for the soft component is seen, indicating
that $kT_{\rm in}\ll 0.5$\,keV, and the photon index is $\Gamma=1.8$. These
values are remarkably similar to the typical hard state behaviour seen,
e.g., in Cyg~X-1. \emph{We interpret the decreases in the ASM countrate as
  evidence for transitions from the normal soft state into the hard state
  of \lmcthree}.

\begin{figure}
\begin{center}
\includegraphics[width=0.5\textwidth]{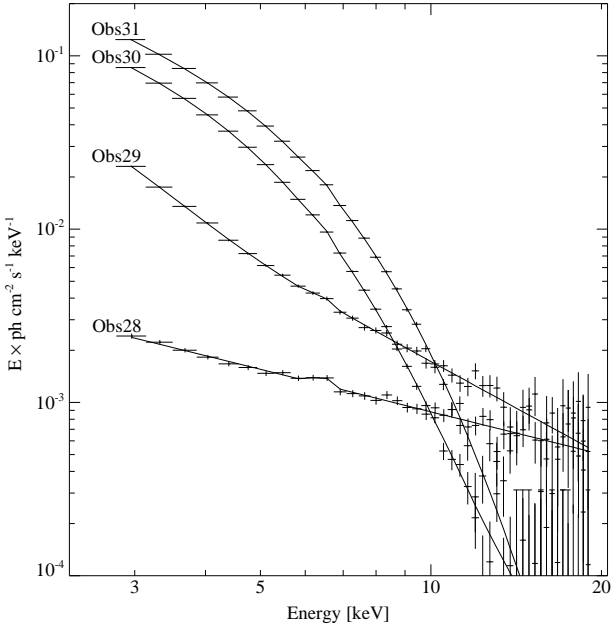}
\end{center}
\caption{Spectral evolution of \lmcthree\ from the hard state of
  Observation~28 to the normal soft state seen in Observations~30
  and~31. Shown are the unfolded photon spectra, multiplied by
  photon energy, the lines denote the best fit
  model.}\label{fig:x3specvar}
\end{figure}

In Fig.~\ref{fig:x3specvar} we display the unfolded photon spectrum of
Obs28, together with the spectra of the three subsequent monitoring
observations, showing the transition from the pure hard state spectrum back
to the soft state.  Obs28 is the only observation of our
campaign in which there is no detectable evidence for the soft component.
The occurance of a pure hard state is thus either a very rare event for
\lmcthree, or, if it is associated with each of the dips observed by the
ASM, it is of relatively short duration. The data presented here represent
the first observational detection of a pure hard state in \lmcthree.  We
speculate, however, that the data taken on \textsl{HEAO~1} day 461 and
mentioned by White \& Marshall \shortcite{white:84a} might have been
observed during one of the soft to hard state transitions. During this
observation, however, the source was too faint for \textsl{HEAO~1} to
obtain even a reliable measurement of the hardness ratio. Furthermore,
during 2000~April another hard state was seen \cite{boyd:00a}, with
the timing behavior being hard state like \cite{homan:00a}.

\begin{figure}
\begin{center}
\includegraphics[width=0.5\textwidth]{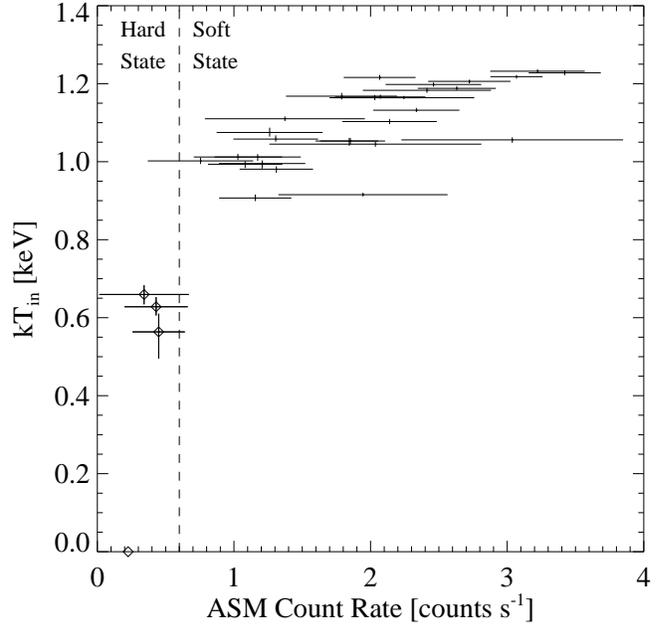}
\end{center}
\caption{The disk blackbody temperature, $kT_{\rm in}$ as a
  function of the 1\,day averaged ASM count rate. Spectra
  observed when the ASM count rate was less than about
  0.6\,counts\,sec$^{-1}$ (indicated by the dashed line) exhibit
  signs for a transition to the hard state.}\label{fig:x3tinasm}
\end{figure}

The hard state observation Obs28 is the observation with the lowest
PCA count rate of our campaign, it was also performed close to one of
the lowest ASM fluxes seen so far. In Fig.~\ref{fig:x3tinasm} we
display $kT_{\rm in}$ as a function of the ASM countrate averaged over
1\,d worth of ASM dwellings centered on the observation. In the figure
we have identified spectra for which $kT_{\rm in}<0.8$\,keV.  All of
these spectra also show a power law component which is relatively
strong compared to the soft spectral component. These data indicate
that for ASM count rates $\aproxlt 0.6$\,cps a transition to the hard
state behaviour is probable. Therefore, it is possible to use the ASM
count rate as a trigger for future systematic studies of the soft to
hard transition in \lmcthree.

Our monitoring observations are too short to be able to characterize
the variability of \lmcthree. We computed the average rms variability
of the source from the background subtracted 16\,s lightcurves.
Although there is an apparent trend of the rms variability to increase
with lower source flux, this trend is fully consistent with systematic
variations due to PCA background model fluctuations at the 2\% level.
This level is consistent with the uncertainty of the background model
found in the time series analysis presented in paper~I, and also with
the adjustments to the background flux found during the spectral
modeling process. It is therefore impossible to check our
interpretation of Obs28 being a hard state or an intermediate
\cite{mendez:97a} state from its temporal characteristics.  Much
longer hard state observations ($>100$\,ksec) are needed for this
determination.

Outside of the hard states, \lmcthree\ exhibits the standard behaviour
of a classical soft state source.  Fig.~\ref{fig:x3cor} displays
several correlations between the spectral fit parameters. For power
law indices $\Gamma\aproxlt 4$ there is a clear correlation between
the power law flux and $\Gamma$: the power law component hardens with
decreasing power law flux (Fig.~\ref{fig:x3cor}a). Such behaviour has
been noted for GX~339$-$4 \cite{wilms:98c} and the Seyfert~1 galaxy
NGC~5548 \cite{chiang:99a}, albeit in their hard spectral states.
Figs.~\ref{fig:x3cor}b and~c verify the results of Ebisawa et al.
\shortcite{ebisawa:93a} that during the soft state there is no strong
correlation between the soft and the hard component. For $\Gamma
\aproxgt 4$, $\Gamma$ does not correlate with the accretion disk
temperature, $kT_{\rm in}$.  There is a slight trend for the power law
flux and $\Gamma$ to increase with $kT_{\rm in}$; however, the
statistics of the data are not good enough to claim any firm
correlation between these two parameters.  In conjunction with the
above spectral results for \lmcthree, in paper~I we noted that the
Fourier frequency-dependent coherence function indicated that the soft
and hard X-ray components were uncorrelated with each other for
\lmcone\ (the variability of \lmcthree\ was too weak to perform such an
analysis).

Finally, we consider the behaviour of the soft spectral component.
The normalization of the multi-temperature disk blackbody, $A_{\rm
  disk}$, is nearly constant throughout the monitoring observations
(Fig.~\ref{fig:x3cor}d), while $kT_{\rm in}$ strongly varies. A
similar behaviour has been found in those soft state black hole
candidates where the luminosity is dominated by the soft component
\cite{tanaka:95a}.  If $A_{\rm disk}$ is interpreted literally, then
$A_{\rm disk}\propto (r_{\rm in}/d)^2 \cos\theta$, where $r_{\rm in}$
is the innermost radius of the accretion disk, $d$ is the distance of
the source, and $\theta$ is the disk inclination angle.  In
Fig.~\ref{fig:x3cor}d we display the values of $r_{\rm in}
\cos^{1/2}\theta$ corresponding to our modeled values of $A_{\rm
  disk}$ on the upper $x$-axis. The parameters for $r_{\rm in}$ found
here are typical for galactic black hole candidates \cite{yaqoob:93b}.
Interpreted in this framework, the constancy of $A_{\rm disk}$ for
\lmcthree\ could imply that the geometrical configuration of the
accretion disk does not appreciably change during the soft state
episodes of the object, and that all changes in the disk luminosity
are accounted for via temperature changes.  We also note that for disk
blackbody temperatures $kT_{\rm in} \aproxlt 0.9$\,keV, the data
approximately agree with the proportionality of $A_{\rm disk} \propto
T_{\rm in}^{-8/3}$ which is expected if $T_{\rm in} \propto r_{\rm
  in}^{-3/4}$ and the inner disk radius moves outward, with relatively
little bolometric luminosity change, during a soft-to-hard state
transition (Fig.~\ref{fig:x3cor}d, dashed line).

The literal interpretation of $A_{\rm disk}$ and $kT_{\rm in}$,
however, is not without problems \cite{merloni:00a}. The innermost
region of thin accretion disks is more complicated than assumed in the
theory underlying the multi-temperature disk blackbody model.  Even in
simple $\alpha$ accretion disk models \cite{shakura:73a}, the inner
region of the accretion disk is dominated by electron scattering. Thus
the emerging spectrum is slightly Comptonized. As a result, the
observed $kT_{\rm in}$ does not correspond to the temperature of the
inner edge of the disk, but rather equals the colour temperature of
the disk \cite{ebisawa:91a,shimura:95a}.  Correcting for this effect
is strongly model dependent, see, e.g., Shimura \& Takahara
\shortcite{shimura:95a} and Merloni et al.  \shortcite{merloni:00a}
for discussions. Furthermore, the presence of a hard spectral
component can also influence $A_{\rm disk}$, as we mentioned above
(see also paper~I and Sect.~\ref{x1:longterm}).  We therefore consider
it dangerous to use correlations between a (corrected) $kT_{\rm in}$
and $r_{\rm in}$ fit parameters to draw firm conclusions on the
geometry or physical environment of the observed system.  The apparent
independency of the disk normalization from its characteristic
(colour) temperature $kT_{\rm in}$, however, is striking in the case
of \lmcthree. Any accretion disk theory attempting to model this
object will have to explain this independency.

Finally, we note that no clear correlation is found between the
equivalent width of the iron line and the other spectral parameters.
This is consistent with our interpretation of the iron line parameters
as upper limits.

\section{LMC X-1: Long Time-Scale Spectral Variability }\label{sec:lmcx1}

\subsection{Introduction}

Like \lmcthree, \lmcone\ was also discovered during the \textsl{UHURU}
scans of the LMC. Its optical companion is a O(7--9)III star with a
mass function of $0.144\,\rm M_\odot$ and an orbital period of 4.2\,d
(Cowley et al., 1995, Hutchings et al. 1987, Hutchings, Crampton \&
Cowley, 1983) \nocite{cowley:95a,hutchings:87a,hutchings:83a} in a
photoionized He II nebula (Bianchi \& Pakull 1985; Pakull \& Angebault
1986).  \nocite{bianchi:85a,pakull:86a} The luminosity of \lmcone\ is
about $2\times 10^{38}$\,erg/s \cite{long:81a} and was found to be
quite constant \cite{sunyaev:90a}.  

The X-ray spectrum of \lmcone\ is similar to that of \lmcthree\ 
\nocite{ebisawa:89a}(Ebisawa, Mitsuda \& Inoue, 1989; paper~I and
references therein), although the relative flux of the disk black body
with respect to the power law is smaller.  There is again evidence for
a weak Fe emission line.

The long term behavior of \lmcone\ during 1996 as seen by RXTE has
been analyzed by Schmidtke, Ponder \& Cowley
\shortcite{schmidtke:99a}. In agreement with previous observations
these authors did not find evidence for any systematic variability on
long timescales. We will, therefore, not re-analyse these data but
concentrate in what follows on our monitoring data from 1997 and 1998.

\subsection{Long Term Variability of LMC X-1}\label{x1:longterm}

\input{table3.tex}
\input{table4.tex}

\begin{figure*}
\begin{center}
\includegraphics[width=0.4\textwidth]{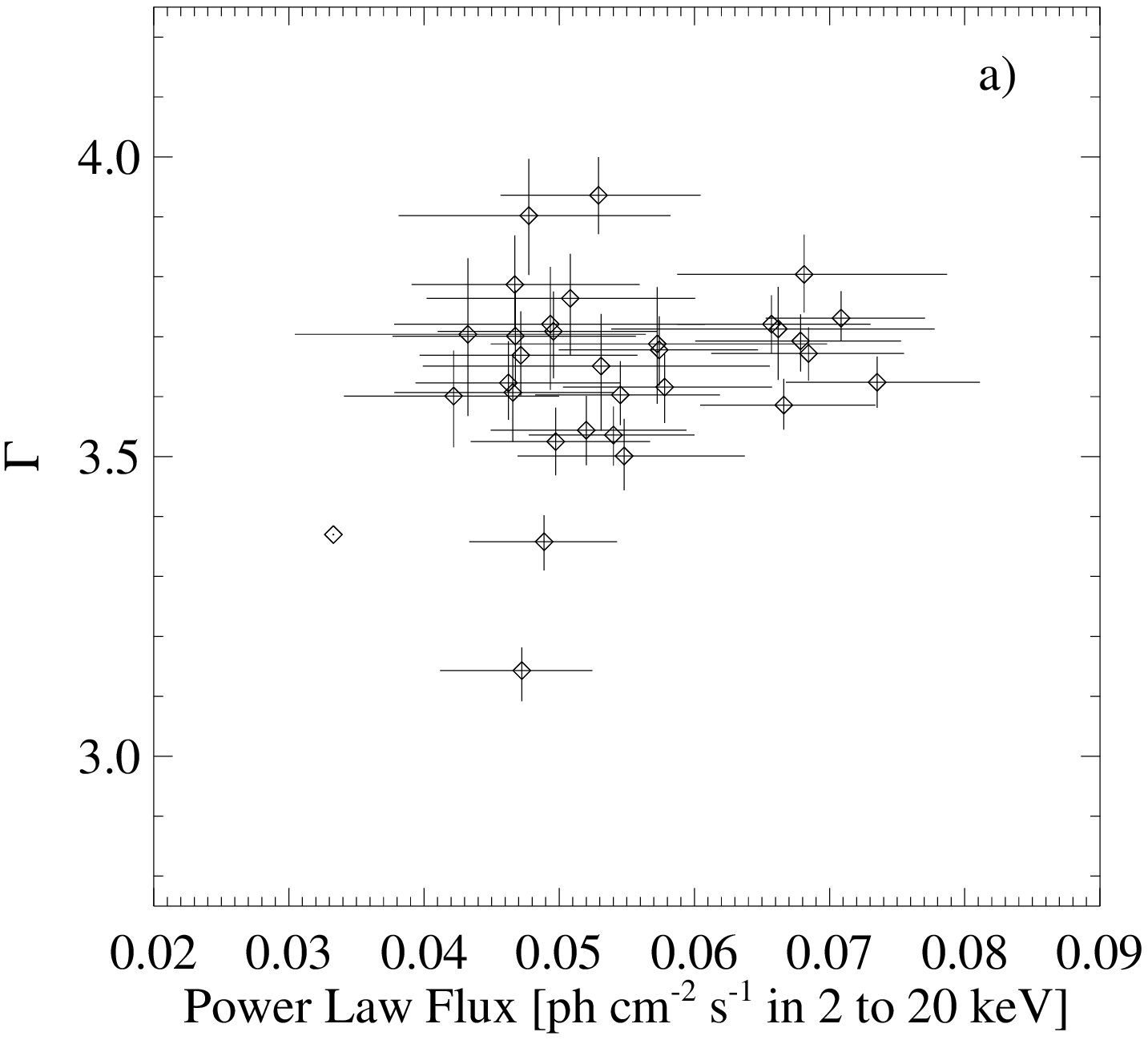}
\includegraphics[width=0.4\textwidth]{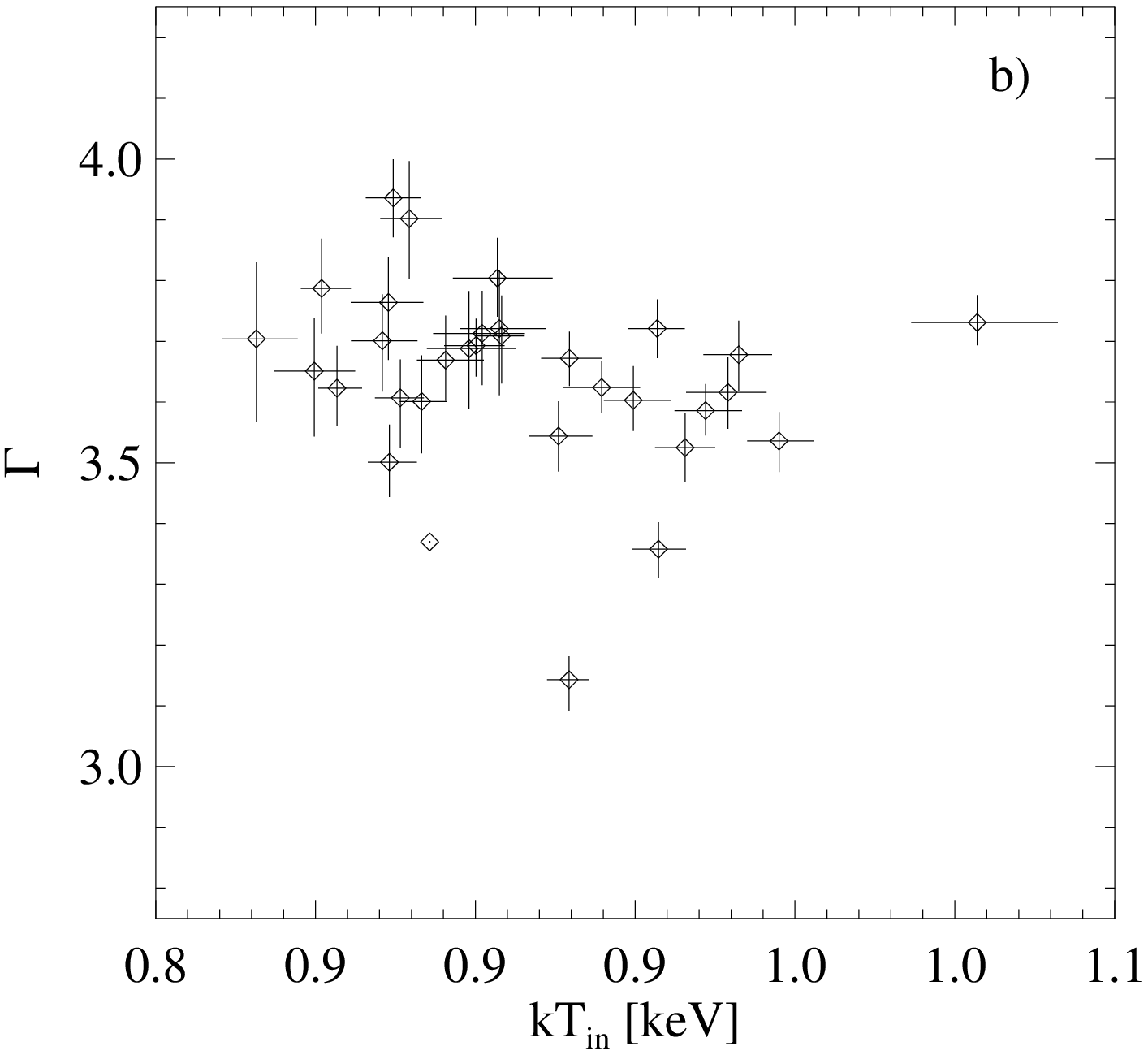}

\includegraphics[width=0.4\textwidth]{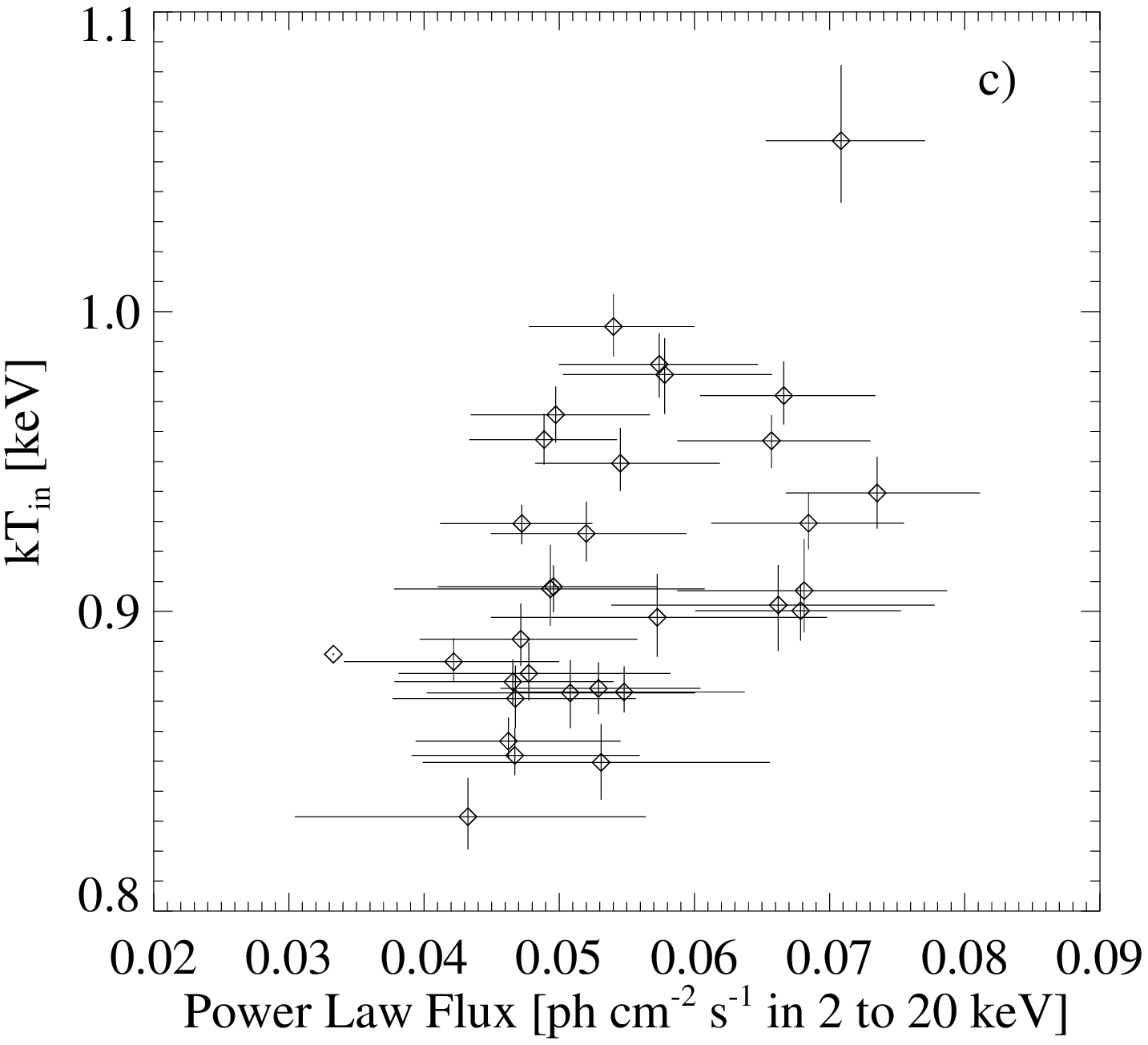}
\includegraphics[width=0.4\textwidth]{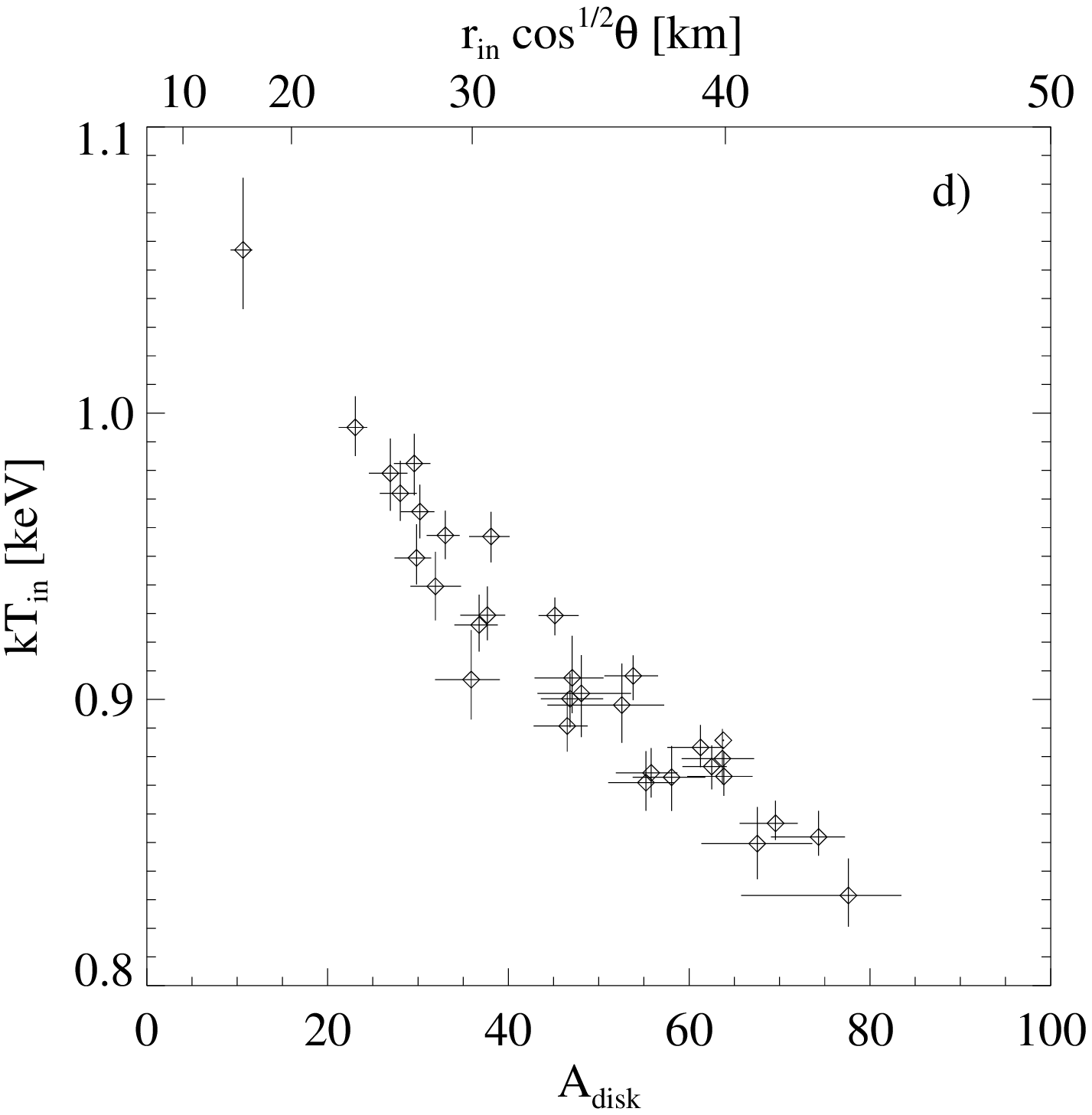}
\end{center}
\caption{Correlation between the spectral parameters for
  \lmcone. Note that the scales of the plots differ from those of
  Fig.~\ref{fig:x3cor}!}\label{fig:x1cor} 
\end{figure*}

The data of the monitoring observations of \lmcone\ were treated
in the same way as those for \lmcthree. A log of the observations
can be found in Tab.~\ref{tab:x1log}. During the campaign, the
count rate of \lmcone\ was on average a factor of $\sim 2$ smaller
than that of \lmcthree. Again, we modeled the PCA data using a
multi-temperature disk blackbody plus a power law and an iron
line. We assumed an equivalent hydrogen column of $N_{\rm
  H}=7.2\times 10^{21}\,\rm cm^{-2}$ (Lister-Staveley, 1999, priv.\ 
comm., see also paper~I), as is appropriate from the radio
measurements. The results of the spectral modeling are shown in
Tab.~\ref{tab:x1fits}. 

As for \lmcthree, the \lmcone\ spectra are well-described by a
phenomenological model consisting of a disk blackbody (with $kT_{\rm
  in}\approx 0.8$--1\,keV) and a soft ($\Gamma \aproxgt 3.5$) power
law component.  Fig.~\ref{fig:x1cor} displays the correlations among
the fit parameters for \lmcone.  The most striking aspect of
Fig.~\ref{fig:x1cor} is the apparent strong correlation between the
phenomenological disk blackbody temperature, $kT_{\rm in}$, and the
disk blackbody normalization, $A_{\rm disk}$.  Furthermore, there is a
weaker correlation between the disk temperature and the 2--20\,keV
power law flux.  These correlations are such that the disk temperature
decreases by $\sim20\%$ while its normalization increases by a factor
of $\sim 4$.  This yields an overall increase of the blackbody flux by
a factor of $\sim 1.6$.  There is at the same time, however, a
decrease by a factor of $\sim 2$ in the power law flux.  In contrast
to \lmcthree\, the contribution of the power law makes up a greater
fraction of the total 2--20\,keV flux in \lmcone; therefore, the
decrease[increase] in disk temperature[normalization] corresponds to
an overall decrease in 2--20\,keV flux by only $\approx 20\%$.

\begin{figure}
\includegraphics[width=0.4\textwidth]{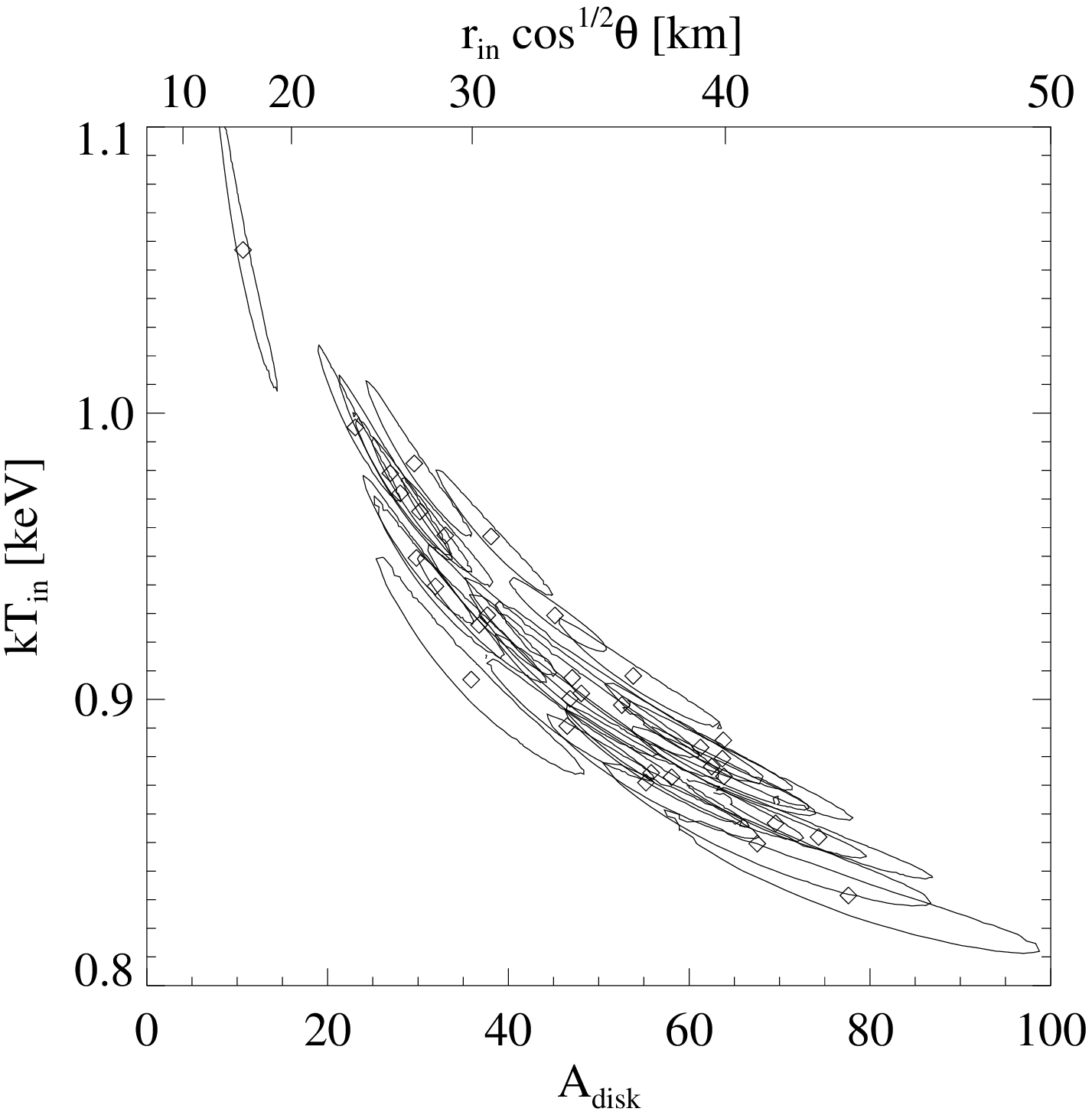}
\caption{Same plot as Fig.~\ref{fig:x1cor}d, but including 99\% error
  contours for each of the fits. The error contours indicate a strong
  \emph{systematic} correlation between $kT_{\rm  in}$ and $A_{\rm
  disk}$.}\label{fig:x1tindcor}
\end{figure}

There are two possibilities to explain these spectral correlations.
First, it is possible that, as in \lmcthree, the overall decrease in
2--20\,keV flux corresponds to the beginnings of a transition to a
low/hard state.  Arguing against this notion, however, is that we do
not see any corresponding hardening of the power law flux.  Second, a
more likely scenario is that we are seeing a complicated systematic
interaction between the phenomenological disk component and power law
component.  For the soft photon indices seen in \lmcone\ there is a
large contribution from the power law in the low-energy PHA channels.
We therefore expect a strong systematic correlation between the power
law flux, the disk temperature, and the disk normalization. In
Fig.~\ref{fig:x1tindcor} we display the 99\% error contours computed
for all spectra from the monitoring campaign.  A strong correlation
between the parameters that is consistent with the general trend seen
in Fig.~\ref{fig:x1cor}d is apparent. This correlation within the
individual datasets strongly points towards our second argument, that
at least part of the correlation is due to systematic errors in our
model. We note, however, that the total variation of the parameters is
much larger than the uncertainty of the individual data points. It is
possible, therefore, that at least part of the $kT_{\rm in}$--$A_{\rm
  disk}$ correlation is indeed real. Observations with instruments
with a lower energy boundary, however, are needed to verify this
correlation.

Apart from a 50\,d periodicity which is seen in many weak ASM sources
(Benlloch-Garc\'\i{}a et al. 2000, in preparation) and is likely due
to the nodal precession of the \textsl{RXTE} orbit, no significant
periodicity was found in the ASM data. The overall ${\cal O}(10\%)$
flux variability exhibited on the long time scales of the ASM and our
monitoring observations is comparable to the variability observed on
time scales of $\aproxlt 1$\,ksec, as discussed in paper~I.  In
paper~I we attributed the variability of \lmcone\ to accretion rate
variations in a wind-fed accretion system.

\section{The Superorbital Period of LMC X-3}\label{sec:softphys}  

The 100--250\,day periodicities seen in the ASM lightcurve of
\lmcthree\ are significantly longer than the measured orbital period
of 1.7\,days.  For the case of \mbox{Her~X-1}, the 35\,day
super-orbital period (Giacconi et al., 1973 \nocite{giacconi:73a}) was
interpreted almost immediately as the result of periodic obscuration
by a precessing accretion disk which is tilted with respect to the
plane of the binary system \cite{katz:73a}.  The large amplitude of
the X-ray flux (a factor of 30 between the ASM count rate during the
``main on'' and the ``low state'' is typical; see, e.g., Coburn et
al., 1999\nocite{coburn:00a}), as well as numerous other observations
(see Scott \& Leahy 1999 \nocite{scott:99a} and references therein),
imply a near edge-on ($i \sim 85^\circ$) disk.  In addition, a
fraction of the central X-ray flux of \mbox{Her~X-1} is likely
scattered into our line of sight by an extended corona (Stelzer et al.
1999 and references therein).  As discussed in the introduction,
possible driving mechanisms for the disk-warp include radiation
pressure due to the luminous central X-ray source
\cite{pringle:96a,maloney:96a,maloney:97a,maloney:98a}, torques
exerted by an accretion disk wind \cite{schandl:94a,schandl:96a}, or
tidal forces (Larwood \nocite{larwood:98a} 1998, and references
therein).

For the case of \lmcthree, if the super-orbital variability were also
due to changes in the accretion disk geometry, we should be able to
observe strong variations in $N_{\rm H}$ as are observed for the
turn-on of the Main-On state of \mbox{Her~X-1} \cite{kuster:99a}.
Were the large intensity drops in \lmcthree\ by a factor of more than
five at 5\,keV seen solely due to variations in $N_{\rm H}$, we would
expect columns on the order of $N_{\rm H} > 5\times 10^{23}\,\rm
cm^{-2}$ blocking our line of sight. Although such columns are readily
detectable with RXTE, they were not observed.  Given the generally
good model fits to the observed data and satisfactory residuals, only
small changes of $N_{\rm H}$ above the interstellar value appear to be
allowed by the data. We conclude that the strong intensity variations
of \lmcthree\ are not due to the accretion disk blocking our line of
sight.

\lmcthree\ is spectrally similar to the X-ray binary \mbox{V1408~Aql}
(\mbox{=4U1957$+$11}), which also shows evidence of long term
(100--300\,day) periodicities \cite{nowak:99b}. Coupled with evidence
from recent optical observations \cite{hakala:99a}, we have argued
that \mbox{V1408~Aql} also exhibits evidence of a warped precessing
disk \cite{nowak:99b}.  The ASM flux of \mbox{V1408~Aql}, however,
varies only by a factor of $\aproxlt 2$ (consistent with warp
inclination angle changes of $\pm 20^\circ$; Nowak \& Wilms
\nocite{nowak:99b} 1999), which is far less than the factor of
$\aproxgt 5$ ASM flux variations seen for \lmcthree.  Such large
amplitude variations would be very difficult to achieve unless the
inclination of the \lmcthree\ system were comparable to that of
\mbox{Her~X-1}\footnote{Van der Klis et al.
  \shortcite{vanderklis:83a}, however, argue for an inclination of
  \lmcthree\ comparable to that which we inferred for
  \mbox{V1408~Aql}, i.e., $i \sim 70^\circ$.  Recent X-ray
  observations of \lmcthree\ (Boyd 1999, priv. comm.) show tentative
  evidence of an orbital modulation. This might indicate some warping
  of at least the outer disk edge (see, for example, Dubus et al.
  \nocite{dubus:99a} 1999), in conjunction with the other long term
  variability mechanisms  discussed below.}.  Given the lack of
any strongly detected variability on the orbital time scale in the ASM
lightcurve of \lmcthree, this seems unlikely.

As discussed in \S\ref{sec:x3var}, spectroscopy yields strong indications
that \lmcthree\ is periodically transitting into and out of the canonical
low/hard X-ray spectral state of galactic black holes. Such transitions are
typical for black holes as they drop below $\sim 5\%$ of their Eddington
luminosity (Nowak \nocite{nowak:95a} 1995; and references therein).  If the
peak ASM flux corresponds to $\sim 30\%~{\rm L_{Edd}}$, then the observed
transition flux threshhold of $\sim 0.6$\,cps in the ASM is roughly
consistent with 5\% of the Eddington luminosity. The question then arises
of why \lmcthree\ apparently shows evidence for large accretion rate
variations, whereas \lmcone\ does not and \mbox{V1408~Aql} might
not\footnote{\mbox{V1408~Aql} historically has had far fewer X-ray
  observations than \lmcthree; therefore, the range of its X-ray spectral
  variations is far less certain.}. 

One possible explanation for the periodic variability seen in
\lmcthree\ is a Compton heated accretion disk wind-driven limit cycle,
as discussed by Shields et al. \shortcite{shields:86a}. At large
radii, in the optically thin upper atmosphere of the outer disk, the
central X-ray source will heat the disk temperature to the inverse
Compton temperature given by
\begin{equation}
4 k T_{\rm IC} \equiv \int F_\nu ~ h \nu ~ {\rm d}\nu / \int F_\nu
{\rm d}\nu ~~~,
\end{equation}
where $F_\nu$ is the central source flux per unit frequency
\cite{begelman:83a,begelman:83b,shields:86a,schandl:94a}. If the sound
speed for the inverse Compton temperature exceeds the escape velocity in
the outer disk, a coronal wind can be driven off of the disk.  For this to
occur, the outer disk must have a radius
\begin{equation}
R_{\rm out} \aproxgt 10^{11} ~{\rm cm} ~ \left ( \frac{3 {\rm keV}}{kT_{\rm
      IC}} \right ) \left ( \frac{{\rm M}}{{\rm M_\odot}} \right ) ~~~.
\end{equation}
\lmcone, which is likely a wind-fed accretion system (paper~I, and
references therein), probably has a much smaller circularization
radius than this value, perhaps substantially so, and therefore is not
expected to be driving off a wind from its outer disk.
\mbox{V1408~Aql}, assuming a $2~{\rm M_\odot}$ primary and a $1~{\rm
  M_\odot}$ companion, has a disk circularization radius of $\sim
4\times 10^{10}$\,cm, and also is not expected to be driving off a
substantial wind (although it may show signs of a photoionized
atmosphere; Nowak \& Wilms \nocite{nowak:99b} 1999). For the measured
system parameters of \lmcthree, the disk circularization radius is
$\sim 1.6 \times 10^{11}$\,cm, and therefore may be driving a
substantial Compton-heated wind.

This opens up the possibility that the central X-ray source of \lmcthree\ 
drives off a sufficiently strong wind such that it ``starves'' itself of
fuel.  The mass accretion rate deficit then will propagate inwards on a
viscous time scale.  The central source X-ray flux will decrease thereby
shutting off the wind, and a limit-cycle can then occur on a viscous time
scale. This time scale can be of order of 100 days if $\alpha$, the usual
disk viscosity parameter, is $\approx 10^{-2}$ \cite{shields:86a}. One
requirement of the instability discussed by Shields et al. is that at the
peak mass flux of the wind, a large fraction, $\aproxgt 90\%$, of the total
accretion flow is expelled from the system.  This is consistent with the
large amplitude variability exhibited by \lmcthree\ in the ASM lightcurve.
As one expects the optical flux to come from large radii and the accretion
rate fluctuations to propagate inwards from the outer disk, this picture is
also consistent with the 20\,day optical lead observed by Cowley et al.
\shortcite{cowley:91a}.

To verify a picture such as discussed by Shields et al.
\shortcite{shields:86a}, one must verify that the low/hard flux
periods of \lmcthree\ truly correspond to a `canonical' low/hard
state.  Additionally, one needs to search for signs of the X-ray
heated wind.  As regards the former possibility, sufficiently long
RXTE observations during the hard state ($> 100$\,ksec) should be able
to detect the characteristically large amplitude variability of this
state.  Furthermore, it recently has been noted that low/hard states
of galactic black holes, as opposed to high/soft states, typically
show strong radio emission (Fender et al. 1999, Corbel et al.  1998,
Zhang et al. 1997\nocite{fender:99b,corbel:98a,zhang:97e} and
references therein).  A previous radio survey of \lmcone\ and
\lmcthree\ at 3.5 and 6.3\,cm did not result in the detection of these
objects. The upper limits for the flux is 1.5\,mJy for \lmcone.  For
\lmcthree, an upper limit of 0.12\,mJy at 3.5\,cm and 0.18\,mJy at
6.3\,cm has been reported \cite{fender:98b}.  These upper limits,
however, were measured during the soft state of \lmcthree.  The above
flux limits do not apply during the recurring episodes of the hard
state. Scaling from Cyg X-1, \lmcthree\ should be observable at a flux
of $\sim 0.05$\,mJy with a sensitive radio telescope during these
phases.

\section{Summary}\label{sec:summary}

In this work, we have presented the analysis of an approximately two
year long \textsl{RXTE} monitoring campaign of \lmcthree\ and \lmcone.
The primary results of these analyses are as follows.

\begin{itemize}

\item Both \lmcone\ and \lmcthree\ can usually be well-described by a
  phenomenological model consisting of a disk blackbody with $kT_{in} \sim
  1$\,keV and a soft ($\Gamma \aproxgt 3$) power law extending down to the
  lowest PHA channels.
  
\item \lmcone\ shows 2--20\,keV energy flux variations with root mean
  square (rms) variability of ${\cal O}(10\%)$ on time scales of weeks to
  months. This is comparable to the rms variability exhibited on time scales
  $\aproxlt 1$\,ksec, as discussed in paper~I.  As elaborated upon in paper
  I, we attributed the variability to accretion rate variations in a
  wind-fed accretion system.
  
\item Accompanying the flux variations in \lmcone\ are correlations
  among the phenomenological disk blackbody temperature, disk
  blackbody normalization, and power law flux.  These correlations
  could be due to variations near a soft state/hard state transition
  region. A strong possibility, however, is that these correlations
  are due to a systematic dependence of the phenomenological fit
  components upon one another.
  
\item \lmcthree\ shows far larger 2--20\,keV flux variations than \lmcone,
  i.e., a factor of ${\cal O}(10)$ luminosity variations on 100--200 day
  time scales. These fluctuations are quasi-periodic, albeit with
  time-varying periods.  At the beginning of the timespan covered by the
  ASM lightcurve, $\sim 100$ and $\sim 180$ day periods are present,
  whereas at the end of the timespan covered by the ASM lightcurve, $\sim
  160$ and $\sim 240$\,day periods are present.
  
\item The flux variations in \lmcthree\ are correlated with clear spectral
  variations.  At high ASM count rates, the flux variations are
  predominantly associated with variations of the phenomenological disk
  blackbody temperature, $kT_{\rm in}$, while maintaining a relatively
  constant disk blackbody normalization, $A_{\rm disk}$. As the ASM count
  rate drops below $\approx 0.6$\,cps, the disk blackbody temperature
  drops, the disk blackbody normalization increases, and the power law
  photon index hardens to $\Gamma \sim 1.8$. 
  
\item The spectral variations in \lmcthree\ are consistent with
  recurring state transitions between the canonical high/soft and
  low/hard X-ray states of galactic black holes. One possible
  mechanism for explaining the observed periodicities and spectral
  variations is a wind driven limit-cycle, as discussed by Shields et
  al.  \shortcite{shields:86a}.

\end{itemize}

As discussed above, there are several ways of determining whether the
low flux/hard spectral states do represent canonical low/hard states
of galactic black holes.  First, a relatively long ($>100$\,ksec) RXTE
observation can determine whether the fast time scale variability is
that expected. Second, one can search for an increase of the radio
flux concurrent with the transition to the low/hard state, as has been
seen for GX~339$-$4 (Fender et al., 1999\nocite{fender:99b}) and Cyg
X-1 (Zhang et al., 1997\nocite{zhang:97e}).  Evidence of a
Compton-heated wind should also be searched for in \lmcthree.
Sufficiently long observations with a sensitive X-ray spectroscopy
instrument may reveal spectral evidence for a wind (e.g., emission
lines in the 1\,keV region) that may be driving a limit cycle in the
\lmcthree\ system.

\section*{Acknowledgements}
The research presented in this paper has been financed by NASA grants
NAG5-3225, NAG5-4621, NAG5-4737, NSF grants AST95-29170, AST98-76887, DFG
grant Sta 173/22, and a travel grant to J.W.  and K.P.  from the Deutscher
Akademischer Austauschdienst.  We thank Didier Barret, Omer Blaes,
W\l{}odzimierz Klu\'zniak, Phil Maloney, Chris Reynolds, Richard
Rothschild, Greg Shields, and Ron Taam for helpful discussions.  Lister
Staveley-Smith provided us with the unpublished $N_{\rm H}$ values for
\mbox{LMC~X-1} and \mbox{LMC~X-3}.  MAN and JW acknowledge the hospitality
of the Aspen Center for Physics and all participants of the Aspen 1999
summer workshop on ``X-ray Probes of Relativistic Astrophysics'' for
discussions.

\end{document}

%% file: table1.tex
\begin{table}
\caption{Observing log of the monitoring observations of LMC X-3.\label{tab:x3log}}
\begin{center}
\begin{tabular}{cccc}
 Obs. &  Date & Exposure &Count Rate \\ 
 & ymd  & sec  & counts\,s$^{-1}$ \\
01 & 1996.12.05 & 139800 &  $ 390.1\pm0.1$ \\ 
\noalign{\vskip 2pt}
02 & 1996.12.29 & ~~ 7800 &  $ ~ 94.3\pm0.1$ \\ 
\noalign{\vskip 2pt}
03 & 1997.01.16 & ~~ 7600 &  $ ~ 83.6\pm0.1$ \\ 
\noalign{\vskip 2pt}
04 & 1997.02.07 & ~~ 9300 &  $ 123.7\pm0.1$ \\ 
\noalign{\vskip 2pt}
05 & 1997.03.09 & ~~ 8000 &  $ 352.2\pm0.2$ \\ 
\noalign{\vskip 2pt}
06 & 1997.03.21 & ~~ 7200 &  $ 374.6\pm0.2$ \\ 
\noalign{\vskip 2pt}
07 & 1997.04.17 & ~~ 8500 &  $ 318.5\pm0.2$ \\ 
\noalign{\vskip 2pt}
08 & 1997.05.05 & ~~ 5200 &  $ 319.3\pm0.3$ \\ 
\noalign{\vskip 2pt}
09 & 1997.05.27 & ~~ 8400 &  $ 289.4\pm0.2$ \\ 
\noalign{\vskip 2pt}
10 & 1997.06.18 & ~~ 8100 &  $ 246.0\pm0.2$ \\ 
\noalign{\vskip 2pt}
11 & 1997.07.04 & ~~ 6200 &  $ 234.8\pm0.2$ \\ 
\noalign{\vskip 2pt}
12 & 1997.08.03 & ~~ 5900 &  $ 148.7\pm0.2$ \\ 
\noalign{\vskip 2pt}
13 & 1997.08.19 & ~~ 8100 &  $ ~ 35.9\pm0.1$ \\ 
\noalign{\vskip 2pt}
14 & 1997.09.09 & ~~ 9400 &  $ 102.6\pm0.1$ \\ 
\noalign{\vskip 2pt}
15 & 1997.09.12 & ~~ 7900 &  $ 113.3\pm0.2$ \\ 
\noalign{\vskip 2pt}
16 & 1997.09.19 & ~~ 8500 &  $ 113.4\pm0.1$ \\ 
\noalign{\vskip 2pt}
17 & 1997.09.23 & ~10300 &  $ 153.7\pm0.1$ \\ 
\noalign{\vskip 2pt}
18 & 1997.10.11 & ~~ 7300 &  $ 127.7\pm0.2$ \\ 
\noalign{\vskip 2pt}
19 & 1997.11.02 & ~~ 8400 &  $ 152.3\pm0.2$ \\ 
\noalign{\vskip 2pt}
20 & 1997.11.23 & ~~ 8900 &  $ 111.5\pm0.1$ \\ 
\noalign{\vskip 2pt}
21 & 1997.12.12 & ~~ 3300 &  $ 166.8\pm0.3$ \\ 
\noalign{\vskip 2pt}
22 & 1998.01.06 & ~~ 8200 &  $ 268.0\pm0.2$ \\ 
\noalign{\vskip 2pt}
23 & 1998.01.24 & ~~ 6100 &  $ 254.0\pm0.2$ \\ 
\noalign{\vskip 2pt}
24 & 1998.02.20 & ~~ 9100 &  $ 213.4\pm0.2$ \\ 
\noalign{\vskip 2pt}
25 & 1998.03.12 & ~~ 6600 &  $ 182.4\pm0.2$ \\ 
\noalign{\vskip 2pt}
26 & 1998.04.06 & ~~ 9200 &  $ 190.0\pm0.2$ \\ 
\noalign{\vskip 2pt}
27 & 1998.05.07 & ~~ 7600 &  $ 169.0\pm0.2$ \\ 
\noalign{\vskip 2pt}
28 & 1998.05.29 & ~~ 9400 &  $ ~~ 9.1\pm0.1$ \\ 
\noalign{\vskip 2pt}
29 & 1998.06.29 & ~~ 3400 &  $ ~ 42.5\pm0.2$ \\ 
\noalign{\vskip 2pt}
30 & 1998.07.20 & ~~ 9800 &  $ 137.3\pm0.1$ \\ 
\noalign{\vskip 2pt}
31 & 1998.08.12 & ~~ 7700 &  $ 133.6\pm0.1$ \\ 
\noalign{\vskip 2pt}
32 & 1998.09.02 & ~~ 4200 &  $ 264.7\pm0.3$ \\ 
\noalign{\vskip 2pt}
33 & 1998.09.30 & ~~ 7600 &  $ ~ 33.2\pm0.1$ \\ 
\noalign{\vskip 2pt}
\end{tabular}\end{center}
{\small Exposure times shown are rounded to the closest
100\,sec. The count rate is background subtracted}
\end{table}

%% file: table2.tex
\begin{table*}
\caption{Results of Spectral Fitting to the LMC X-3 Data.\label{tab:x3fits}}
\begin{center}
\begin{tabular}{clllllcr}
Obs. & \multicolumn{1}{c}{$kT_{\rm in}$} & \multicolumn{1}{c}{$A_{\rm disk}$} &\multicolumn{1}{c}{$\Gamma$} & \multicolumn{1}{c}{$A_{\rm PL}$} & \multicolumn{1}{c}{$A_{\rm Line}$} & \multicolumn{1}{c}{EW} & \multicolumn{1}{c}{$\chi^2/\rm dof$} \\
 & \multicolumn{1}{c}{keV} &  & &\multicolumn{1}{c}{$10^{-1}$} & \multicolumn{1}{c}{$10^{-5}$} & \multicolumn{1}{c}{eV} & \\
01 & $  1.23\pm 0.01$ & $  33.9\pm  0.7$ & $  2.4\pm 0.2$ & $ ~ 0.7^{+ 0.3}_{-  0.2}$ & $ ~  8.5^{+ 11.3}_{-  8.5}$ & $  14 $ & $ 59.5/  47$ \\ \noalign{\vskip 2pt}
02 & $  0.92\pm 0.00$ & $  35.9^{+  0.8}_{-  1.1}$ & $  2.8\pm 0.2$ & $ ~ 0.5\pm 0.2$ & $ ~ 8.4\pm 2.2$ & $  88 $ & $ 72.7/  42$ \\ \noalign{\vskip 2pt}
03 & $  0.91\pm 0.01$ & $  29.7^{+  1.6}_{-  1.7}$ & $  3.0\pm 0.1$ & $ ~ 1.4\pm 0.3$ & $ ~  6.7^{+  2.0}_{-  2.7}$ & $  65 $ & $ 60.8/  42$ \\ \noalign{\vskip 2pt}
04 & $  1.01\pm 0.01$ & $  29.3^{+  0.5}_{-  0.8}$ & $  4.3\pm 0.3$ & $ ~ 3.6^{+ 1.8}_{-  1.0}$ & $ ~ 6.8\pm 2.4$ & $  51 $ & $ 19.3/  41$ \\ \noalign{\vskip 2pt}
05 & $  1.21^{+ 0.00}_{- 0.01}$ & $  29.9^{+  0.6}_{-  0.7}$ & $  3.1\pm 0.1$ & $ ~ 3.4^{+ 0.5}_{-  0.4}$ & $  11.6^{+  4.1}_{-  3.8}$ & $  22 $ & $ 62.6/  41$ \\ \noalign{\vskip 2pt}
06 & $  1.22^{+ 0.00}_{- 0.01}$ & $  31.8^{+  0.4}_{-  0.6}$ & $  3.2\pm 0.1$ & $ ~ 3.3\pm 0.5$ & $  20.5^{+  4.1}_{-  4.5}$ & $  36 $ & $ 59.5/  41$ \\ \noalign{\vskip 2pt}
07 & $  1.23^{+ 0.01}_{- 0.00}$ & $  27.5\pm  0.4$ & $  4.1\pm 0.2$ & $ ~ 5.3^{+ 1.5}_{-  1.1}$ & $ ~  7.5^{+  4.1}_{-  3.4}$ & $  16 $ & $ 73.3/  41$ \\ \noalign{\vskip 2pt}
08 & $  1.22\pm 0.01$ & $  29.5\pm  0.5$ & $  3.7\pm 0.2$ & $ ~ 3.8^{+ 1.2}_{-  0.9}$ & $  12.0^{+  5.0}_{-  4.7}$ & $  25 $ & $ 46.4/  41$ \\ \noalign{\vskip 2pt}
09 & $  1.20^{+ 0.00}_{- 0.01}$ & $  29.0\pm  0.4$ & $  4.1\pm 0.2$ & $ ~ 4.8^{+ 1.4}_{-  1.1}$ & $  13.7^{+  3.5}_{-  3.7}$ & $  32 $ & $ 49.9/  41$ \\ \noalign{\vskip 2pt}
10 & $  1.17^{+ 0.01}_{- 0.00}$ & $  28.0^{+  0.4}_{-  0.5}$ & $  4.2^{+ 0.3}_{-  0.2}$ & $ ~ 5.5^{+ 1.8}_{-  1.3}$ & $ ~  5.8^{+  3.7}_{-  3.2}$ & $  17 $ & $ 58.3/  41$ \\ \noalign{\vskip 2pt}
11 & $  1.16\pm 0.01$ & $  26.9\pm  0.5$ & $  4.1^{+ 0.3}_{-  0.2}$ & $ ~ 5.3^{+ 1.8}_{-  1.3}$ & $ 10.1\pm 3.9$ & $  31 $ & $ 58.1/  41$ \\ \noalign{\vskip 2pt}
12 & $  1.00\pm 0.01$ & $  29.6^{+  1.2}_{-  1.6}$ & $  3.1\pm 0.1$ & $ ~ 2.1^{+ 0.6}_{-  0.5}$ & $ ~  8.2^{+  2.9}_{-  3.1}$ & $  46 $ & $ 97.1/  41$ \\ \noalign{\vskip 2pt}
13 & $  0.63\pm 0.02$ & $  54.4^{+ 10.7}_{-  8.7}$ & $  2.8\pm 0.1$ & $ ~ 0.7\pm 0.1$ & $ ~  0.0^{+  1.7}_{-  0.0}$ & $ ~  0 $ & $ 29.0/  41$ \\ \noalign{\vskip 2pt}
14 & $  0.98\pm 0.01$ & $  28.6^{+  0.9}_{-  0.7}$ & $  4.6^{+ 0.4}_{-  0.3}$ & $ ~ 4.6^{+ 2.4}_{-  1.2}$ & $ ~  7.3^{+  2.0}_{-  2.5}$ & $  71 $ & $ 34.4/  41$ \\ \noalign{\vskip 2pt}
15 & $  1.00\pm 0.01$ & $  29.4\pm  0.9$ & $  4.2\pm 0.3$ & $ ~ 3.0^{+ 1.6}_{-  0.7}$ & $ ~  6.7^{+  2.4}_{-  2.7}$ & $  56 $ & $ 24.1/  41$ \\ \noalign{\vskip 2pt}
16 & $  1.05\pm 0.01$ & $  24.7^{+  1.4}_{-  1.1}$ & $  4.6\pm 0.3$ & $ ~ 6.9^{+ 2.9}_{-  1.9}$ & $ ~  5.3^{+  2.5}_{-  2.4}$ & $  35 $ & $ 70.2/  41$ \\ \noalign{\vskip 2pt}
17 & $  1.06\pm 0.01$ & $  27.1^{+  0.8}_{-  1.4}$ & $  4.0\pm 0.2$ & $ ~ 4.6^{+ 1.1}_{-  1.0}$ & $ ~  2.5^{+  2.6}_{-  2.5}$ & $  14 $ & $ 75.8/  41$ \\ \noalign{\vskip 2pt}
18 & $  1.01\pm 0.01$ & $  29.6^{+  2.2}_{-  1.0}$ & $  4.2^{+ 0.3}_{-  0.2}$ & $ ~ 4.1^{+ 1.6}_{-  1.1}$ & $ ~  6.4^{+  2.6}_{-  3.0}$ & $  46 $ & $ 26.5/  41$ \\ \noalign{\vskip 2pt}
19 & $  1.06\pm 0.01$ & $  28.6^{+  1.0}_{-  0.7}$ & $  4.4^{+ 0.3}_{-  0.2}$ & $ ~ 5.5^{+ 1.9}_{-  1.3}$ & $ ~  7.5^{+  2.5}_{-  3.0}$ & $  42 $ & $ 42.6/  41$ \\ \noalign{\vskip 2pt}
20 & $  0.99\pm 0.01$ & $  27.8^{+  0.5}_{-  0.8}$ & $  4.3^{+ 0.3}_{-  0.2}$ & $ ~ 4.5^{+ 1.8}_{-  1.1}$ & $ ~  8.0^{+  2.2}_{-  2.7}$ & $  70 $ & $ 17.2/  41$ \\ \noalign{\vskip 2pt}
21 & $  1.08\pm 0.01$ & $  28.3^{+  2.3}_{-  0.8}$ & $  4.3^{+ 0.5}_{-  0.3}$ & $ ~ 4.9^{+ 3.3}_{-  1.5}$ & $  11.9^{+  4.7}_{-  4.6}$ & $  59 $ & $ 40.3/  41$ \\ \noalign{\vskip 2pt}
22 & $  1.19^{+ 0.01}_{- 0.00}$ & $  27.6\pm  0.4$ & $  4.3\pm 0.2$ & $ ~ 7.4^{+ 1.9}_{-  1.4}$ & $ 11.7\pm 3.6$ & $  31 $ & $ 43.8/  41$ \\ \noalign{\vskip 2pt}
23 & $  1.18\pm 0.01$ & $  27.3\pm  0.5$ & $  4.2\pm 0.2$ & $ ~ 7.4^{+ 2.1}_{-  1.6}$ & $  11.2^{+  3.9}_{-  4.5}$ & $  30 $ & $ 38.4/  41$ \\ \noalign{\vskip 2pt}
24 & $  1.13^{+ 0.01}_{- 0.00}$ & $  28.1\pm  0.4$ & $  4.4^{+ 0.3}_{-  0.2}$ & $ ~ 7.8^{+ 2.3}_{-  1.6}$ & $  10.6^{+  3.3}_{-  2.9}$ & $  38 $ & $ 42.3/  41$ \\ \noalign{\vskip 2pt}
25 & $  1.10\pm 0.01$ & $  27.3^{+  0.3}_{-  0.6}$ & $  4.6^{+ 0.4}_{-  0.3}$ & $ ~ 8.4^{+ 3.3}_{-  2.0}$ & $ 10.4\pm 3.4$ & $  45 $ & $ 35.2/  41$ \\ \noalign{\vskip 2pt}
26 & $  1.11\pm 0.01$ & $  27.5^{+  0.5}_{-  0.6}$ & $  4.3\pm 0.2$ & $ ~ 6.5^{+ 1.7}_{-  1.4}$ & $ ~  9.3^{+  3.0}_{-  2.9}$ & $  38 $ & $ 55.8/  41$ \\ \noalign{\vskip 2pt}
27 & $  1.05\pm 0.01$ & $  25.0^{+  1.1}_{-  1.2}$ & $  3.5\pm 0.1$ & $ ~ 4.4^{+ 0.7}_{-  0.6}$ & $ ~ 5.8\pm 3.1$ & $  28 $ & $101.5/  41$ \\ \noalign{\vskip 2pt}
28 &   
&    
& $  1.8\pm 0.0$ & $ ~ 0.1\pm 0.0$ & $ ~  1.2^{+  1.1}_{-  1.2}$ & $  60 $ & $ 26.5/  43$ \\ \noalign{\vskip 2pt}
29 & $  0.56^{+ 0.05}_{- 0.07}$ & $  80.1^{+ 80.5}_{- 30.8}$ & $  2.8\pm 0.1$ & $ ~ 1.1^{+ 0.2}_{-  0.3}$ & $ ~  2.0^{+  2.8}_{-  2.0}$ & $  33 $ & $ 22.7/  41$ \\ \noalign{\vskip 2pt}
30 & $  1.04\pm 0.01$ & $  27.0\pm  0.6$ & $  4.5^{+ 0.3}_{-  0.2}$ & $ ~ 6.3^{+ 2.0}_{-  1.5}$ & $ ~  4.7^{+  2.6}_{-  2.5}$ & $  30 $ & $ 38.1/  41$ \\ \noalign{\vskip 2pt}
31 & $  1.16\pm 0.00$ & $  24.8^{+  0.6}_{-  0.5}$ & $  5.5^{+ 0.8}_{-  0.5}$ & $ 27.7^{+18.7}_{- 10.5}$ & $ ~  7.0^{+  3.9}_{-  3.2}$ & $  24 $ & $ 87.0/  41$ \\ \noalign{\vskip 2pt}
32 & $  1.17\pm 0.01$ & $  27.2^{+  0.8}_{-  0.9}$ & $  3.8\pm 0.1$ & $ ~ 7.7^{+ 1.4}_{-  1.2}$ & $  11.4^{+  5.2}_{-  4.9}$ & $  30 $ & $ 43.3/  41$ \\ \noalign{\vskip 2pt}
33 & $  0.66^{+ 0.02}_{- 0.03}$ & $  57.0^{+ 10.7}_{-  7.9}$ & $  2.7\pm 0.1$ & $ ~ 1.0\pm 0.2$ & $ ~  4.9^{+  2.3}_{-  2.6}$ & $  71 $ & $ 30.3/  41$ \\ \noalign{\vskip 2pt}
\end{tabular}
\end{center}
{\small 
$T_{\rm in}$, $A_{\rm disk}$: peak multi-temperature disk
temperature and normalization.
$\Gamma$: photon index of the power law
$A_{\rm PL}$: Power law normalization 
(photons\,keV$^{-1}$\,cm$^{-2}$\,s$^{-1}$ at 1\,keV).
$A_{\rm Line}$: Line normalization
(photons\,cm$^{-2}$\,s$^{-1}$ in the line), the Gaussian line
was fixed at 6.4\,keV with a width $\sigma$ of 0.1\,keV. 
EW: line equivalent width.
Uncertainties are at the 90\% confidence level for one interesting
parameter ($\Delta \chi^2 = 2.71$), the interstellar equivalent
column was fixed at $N_{\rm H}=3.2\times 10^{20}\,\rm cm^{-2}$. 
}
\end{table*}

%% file: table3.tex
\begin{table}
\caption{Observing log of the monitoring observations of LMC X-1.\label{tab:x1log}}
\begin{center}
\begin{tabular}{cccc}
 Obs. &  Date & Exposure &Count Rate \\ 
 & ymd  & sec  & counts\,s$^{-1}$ \\
01 & 1996.12.06 &  128900 &  $ 151.6\pm0.0$ \\ 
\noalign{\vskip 2pt}
02 & 1996.12.30 & ~~ 8600 &  $ 146.7\pm0.2$ \\ 
\noalign{\vskip 2pt}
03 & 1997.01.18 & ~~ 8200 &  $ 180.4\pm0.2$ \\ 
\noalign{\vskip 2pt}
04 & 1997.02.08 & ~~ 9300 &  $ 149.4\pm0.1$ \\ 
\noalign{\vskip 2pt}
05 & 1997.03.09 & ~~ 3800 &  $ 157.0\pm0.2$ \\ 
\noalign{\vskip 2pt}
06 & 1997.03.21 & ~~ 8200 &  $ 160.8\pm0.2$ \\ 
\noalign{\vskip 2pt}
07 & 1997.04.16 & ~10600 &  $ 163.4\pm0.1$ \\ 
\noalign{\vskip 2pt}
08 & 1997.05.07 & ~~ 8600 &  $ 163.6\pm0.2$ \\ 
\noalign{\vskip 2pt}
09 & 1997.05.28 & ~~ 7000 &  $ 147.2\pm0.2$ \\ 
\noalign{\vskip 2pt}
10 & 1997.06.18 & ~~ 3600 &  $ 161.9\pm0.2$ \\ 
\noalign{\vskip 2pt}
11 & 1997.07.04 & ~~ 3400 &  $ 148.6\pm0.3$ \\ 
\noalign{\vskip 2pt}
12 & 1997.08.01 & ~~ 8500 &  $ 157.1\pm0.2$ \\ 
\noalign{\vskip 2pt}
13 & 1997.08.20 & ~~ 7600 &  $ 171.1\pm0.2$ \\ 
\noalign{\vskip 2pt}
14 & 1997.09.09 & ~~ 9600 &  $ 138.9\pm0.1$ \\ 
\noalign{\vskip 2pt}
15 & 1997.09.12 & ~~ 8300 &  $ 163.3\pm0.2$ \\ 
\noalign{\vskip 2pt}
16 & 1997.09.19 & ~~ 9100 &  $ 120.8\pm0.1$ \\ 
\noalign{\vskip 2pt}
17 & 1997.10.10 & ~~ 9100 &  $ 154.3\pm0.2$ \\ 
\noalign{\vskip 2pt}
18 & 1997.11.01 & ~~ 8800 &  $ 157.0\pm0.2$ \\ 
\noalign{\vskip 2pt}
19 & 1997.11.23 & ~~ 6400 &  $ 156.1\pm0.2$ \\ 
\noalign{\vskip 2pt}
20 & 1997.12.12 & ~~ 5100 &  $ 160.7\pm0.2$ \\ 
\noalign{\vskip 2pt}
21 & 1998.01.04 & ~~ 8700 &  $ 189.4\pm0.2$ \\ 
\noalign{\vskip 2pt}
22 & 1998.01.25 & ~~ 7100 &  $ 168.0\pm0.2$ \\ 
\noalign{\vskip 2pt}
23 & 1998.02.20 & ~~ 9200 &  $ 177.1\pm0.2$ \\ 
\noalign{\vskip 2pt}
24 & 1998.03.12 & ~~ 3300 &  $ 173.9\pm0.3$ \\ 
\noalign{\vskip 2pt}
25 & 1998.04.07 & ~~ 9400 &  $ 178.1\pm0.2$ \\ 
\noalign{\vskip 2pt}
26 & 1998.05.06 & ~~ 5500 &  $ 154.0\pm0.2$ \\ 
\noalign{\vskip 2pt}
27 & 1998.05.28 & ~~ 8900 &  $ 179.1\pm0.2$ \\ 
\noalign{\vskip 2pt}
28 & 1998.06.28 & ~~ 3700 &  $ 178.1\pm0.3$ \\ 
\noalign{\vskip 2pt}
29 & 1998.07.19 & ~~ 9300 &  $ 151.4\pm0.2$ \\ 
\noalign{\vskip 2pt}
30 & 1998.08.13 & ~~ 8000 &  $ 181.2\pm0.2$ \\ 
\noalign{\vskip 2pt}
31 & 1998.09.02 & ~~ 5200 &  $ 143.9\pm0.2$ \\ 
\noalign{\vskip 2pt}
32 & 1998.09.29 & ~~ 8600 &  $ 112.7\pm0.1$ \\ 
\noalign{\vskip 2pt}
\end{tabular}\end{center}
{\small Exposure times shown are rounded to the closest
100\,sec. The count rate is background subtracted}
\end{table}

%% file: table4.tex
\begin{table*}
\caption{Results of Spectral Fitting to the LMC X-1 Data.\label{tab:x1fits}}
\begin{center}
\begin{tabular}{clllllcr}
Obs. & \multicolumn{1}{c}{$kT_{\rm in}$} & \multicolumn{1}{c}{$A_{\rm disk}$} &\multicolumn{1}{c}{$\Gamma$} & \multicolumn{1}{c}{$A_{\rm PL}$} & \multicolumn{1}{c}{$A_{\rm Line}$} & \multicolumn{1}{c}{EW} & \multicolumn{1}{c}{$\chi^2/\rm dof$} \\
 & \multicolumn{1}{c}{keV} &  & &\multicolumn{1}{c}{$10^{-1}$} & \multicolumn{1}{c}{$10^{-5}$} & \multicolumn{1}{c}{eV} & \\
01 & $  0.89\pm 0.00$ & $  63.8\pm  0.0$ & $  3.4\pm 0.0$ & $ ~  4.1\pm 0.0$ & $ 10.7\pm 0.0$ & $  63 $ & $ 80.8/  41$ \\  \noalign{\vskip 2pt}
02 & $  0.96\pm 0.01$ & $  33.0^{+  1.6}_{-  2.1}$ & $  3.4\pm 0.0$ & $ ~  5.9\pm 0.6$ & $  10.8^{+  3.0}_{-  2.8}$ & $  53 $ & $ 84.9/  42$ \\  \noalign{\vskip 2pt}
03 & $  0.93\pm 0.01$ & $  45.2^{+  2.6}_{-  1.8}$ & $  3.1^{+ 0.0}_{-  0.1}$ & $ ~  4.5\pm 0.5$ & $  19.2^{+  3.0}_{-  3.1}$ & $  85 $ & $ 88.2/  42$ \\  \noalign{\vskip 2pt}
04 & $  0.87\pm 0.01$ & $  63.8^{+  3.2}_{-  4.1}$ & $  3.5\pm 0.1$ & $ ~  7.8^{+ 1.2}_{-  1.0}$ & $  13.1^{+  3.0}_{-  2.9}$ & $  67 $ & $ 50.3/  41$ \\  \noalign{\vskip 2pt}
05 & $  0.91\pm 0.01$ & $  47.1^{+  3.5}_{-  4.2}$ & $  3.7\pm 0.1$ & $ ~  8.9\pm 1.8$ & $ ~   8.1^{+  4.3}_{-  4.0}$ & $  47 $ & $ 37.5/  41$ \\  \noalign{\vskip 2pt}
06 & $  0.88\pm 0.01$ & $  61.3^{+  2.6}_{-  3.7}$ & $  3.6\pm 0.1$ & $ ~  6.7\pm 1.1$ & $ ~   8.6^{+  3.1}_{-  2.7}$ & $  51 $ & $ 45.5/  41$ \\  \noalign{\vskip 2pt}
07 & $  0.86\pm 0.01$ & $  69.6^{+  2.5}_{-  4.0}$ & $  3.6\pm 0.1$ & $ ~  7.5^{+ 1.2}_{-  1.0}$ & $  11.6^{+  2.6}_{-  2.5}$ & $  71 $ & $ 95.8/  41$ \\  \noalign{\vskip 2pt}
08 & $  0.88\pm 0.01$ & $  62.5^{+  1.7}_{-  3.2}$ & $  3.6\pm 0.1$ & $ ~  7.4^{+ 1.1}_{-  1.2}$ & $  11.3^{+  2.7}_{-  2.9}$ & $  65 $ & $ 48.7/  41$ \\  \noalign{\vskip 2pt}
09 & $  0.87\pm 0.01$ & $  55.2^{+  3.1}_{-  4.2}$ & $  3.7\pm 0.1$ & $ ~  8.2\pm 1.4$ & $  11.7^{+  2.8}_{-  2.9}$ & $  76 $ & $ 52.7/  41$ \\  \noalign{\vskip 2pt}
10 & $  0.85\pm 0.01$ & $  67.5^{+  6.1}_{-  6.2}$ & $  3.7\pm 0.1$ & $ ~  8.9^{+ 1.9}_{-  2.0}$ & $  13.1^{+  4.0}_{-  4.4}$ & $  78 $ & $ 32.4/  41$ \\  \noalign{\vskip 2pt}
11 & $  0.83\pm 0.01$ & $  77.6^{+  5.9}_{- 11.9}$ & $  3.7\pm 0.1$ & $ ~  7.6^{+ 2.1}_{-  2.0}$ & $ ~  6.6\pm 4.2$ & $  46 $ & $ 37.6/  41$ \\  \noalign{\vskip 2pt}
12 & $  0.93\pm 0.01$ & $  36.8^{+  2.1}_{-  2.7}$ & $  3.5\pm 0.1$ & $ ~  7.7^{+ 1.0}_{-  0.9}$ & $  10.2^{+  2.8}_{-  3.0}$ & $  56 $ & $ 91.2/  41$ \\  \noalign{\vskip 2pt}
13 & $  0.91\pm 0.01$ & $  53.8^{+  2.7}_{-  3.2}$ & $  3.7\pm 0.1$ & $ ~  8.8^{+ 1.2}_{-  1.3}$ & $ ~   9.3^{+  3.1}_{-  3.2}$ & $  50 $ & $ 35.4/  41$ \\  \noalign{\vskip 2pt}
14 & $  0.95\pm 0.01$ & $  29.8^{+  1.6}_{-  2.4}$ & $  3.6\pm 0.1$ & $ ~  8.6^{+ 1.0}_{-  0.9}$ & $ ~   9.4^{+  2.9}_{-  2.8}$ & $  51 $ & $ 54.3/  41$ \\  \noalign{\vskip 2pt}
15 & $  0.85\pm 0.01$ & $  74.3^{+  2.9}_{-  5.3}$ & $  3.8\pm 0.1$ & $ ~  9.0^{+ 1.6}_{-  1.3}$ & $ ~   8.6^{+  2.9}_{-  2.6}$ & $  55 $ & $ 85.9/  41$ \\  \noalign{\vskip 2pt}
16 & $  0.89\pm 0.01$ & $  46.5^{+  2.3}_{-  3.7}$ & $  3.7\pm 0.1$ & $ ~  8.0^{+ 1.3}_{-  1.1}$ & $  11.5^{+  2.8}_{-  2.7}$ & $  73 $ & $ 68.3/  41$ \\  \noalign{\vskip 2pt}
17 & $  1.00\pm 0.01$ & $  23.1^{+  1.3}_{-  1.8}$ & $  3.5^{+ 0.0}_{-  0.1}$ & $ ~  8.0\pm 0.8$ & $  10.5^{+  3.0}_{-  2.7}$ & $  54 $ & $ 34.5/  41$ \\  \noalign{\vskip 2pt}
18 & $  0.97\pm 0.01$ & $  30.2^{+  1.6}_{-  2.1}$ & $  3.5\pm 0.1$ & $ ~  7.3^{+ 0.9}_{-  0.8}$ & $  11.0^{+  2.7}_{-  3.1}$ & $  58 $ & $ 29.9/  41$ \\  \noalign{\vskip 2pt}
19 & $  0.98\pm 0.01$ & $  26.9^{+  1.9}_{-  2.4}$ & $  3.6\pm 0.1$ & $ ~  9.3\pm 1.1$ & $  12.5^{+  3.3}_{-  3.7}$ & $  63 $ & $ 47.0/  41$ \\  \noalign{\vskip 2pt}
20 & $  0.88\pm 0.01$ & $  63.7^{+  3.5}_{-  4.5}$ & $  3.9\pm 0.1$ & $ 10.4^{+ 2.0}_{-  1.8}$ & $ ~   7.2^{+  3.4}_{-  3.5}$ & $  45 $ & $ 36.9/  41$ \\  \noalign{\vskip 2pt}
21 & $  0.96\pm 0.01$ & $  38.1^{+  2.1}_{-  2.4}$ & $  3.7\pm 0.0$ & $ 11.8^{+ 1.2}_{-  1.1}$ & $  13.1^{+  2.9}_{-  3.2}$ & $  59 $ & $ 66.0/  41$ \\  \noalign{\vskip 2pt}
22 & $  0.98\pm 0.01$ & $  29.6^{+  1.8}_{-  2.2}$ & $  3.7\pm 0.1$ & $ ~  9.9\pm 1.1$ & $  10.9^{+  3.3}_{-  3.4}$ & $  53 $ & $ 38.6/  41$ \\  \noalign{\vskip 2pt}
23 & $  0.97\pm 0.01$ & $  28.0^{+  1.8}_{-  2.3}$ & $  3.6\pm 0.0$ & $ 10.4^{+ 0.9}_{-  0.8}$ & $ 14.5\pm 3.0$ & $  66 $ & $ 90.9/  41$ \\  \noalign{\vskip 2pt}
24 & $  0.90\pm 0.01$ & $  52.5^{+  4.7}_{-  8.2}$ & $  3.7\pm 0.1$ & $ ~  9.9\pm 1.9$ & $ ~   7.7^{+  4.5}_{-  5.2}$ & $  40 $ & $ 39.1/  41$ \\  \noalign{\vskip 2pt}
25 & $  0.93\pm 0.01$ & $  37.7^{+  2.0}_{-  3.0}$ & $  3.7\pm 0.0$ & $ 11.7\pm 1.1$ & $  10.7^{+  3.2}_{-  2.7}$ & $  52 $ & $ 40.9/  41$ \\  \noalign{\vskip 2pt}
26 & $  0.87\pm 0.01$ & $  58.1^{+  3.7}_{-  4.3}$ & $  3.8\pm 0.1$ & $ ~  9.6^{+ 1.6}_{-  1.8}$ & $  11.8^{+  3.5}_{-  3.6}$ & $  73 $ & $ 46.4/  41$ \\  \noalign{\vskip 2pt}
27 & $  0.90\pm 0.01$ & $  46.8^{+  3.6}_{-  3.2}$ & $  3.7^{+ 0.0}_{-  0.1}$ & $ 11.8\pm 1.2$ & $  15.9^{+  2.9}_{-  3.0}$ & $  79 $ & $ 44.2/  41$ \\  \noalign{\vskip 2pt}
28 & $  0.90^{+ 0.01}_{- 0.02}$ & $  48.1^{+  5.5}_{-  4.9}$ & $  3.7\pm 0.1$ & $ 11.8\pm 1.9$ & $  14.9^{+  4.5}_{-  4.7}$ & $  75 $ & $ 59.5/  41$ \\  \noalign{\vskip 2pt}
29 & $  0.87\pm 0.01$ & $  55.8^{+  2.9}_{-  3.9}$ & $  3.9\pm 0.1$ & $ 11.9^{+ 1.5}_{-  1.4}$ & $ ~   8.1^{+  2.6}_{-  2.8}$ & $  54 $ & $ 34.5/  41$ \\  \noalign{\vskip 2pt}
30 & $  0.94\pm 0.01$ & $  31.9\pm  2.8$ & $  3.6\pm 0.0$ & $ 11.9^{+ 1.1}_{-  0.9}$ & $  16.9^{+  3.0}_{-  3.5}$ & $  78 $ & $110.5/  41$ \\  \noalign{\vskip 2pt}
31 & $  0.91^{+ 0.02}_{- 0.01}$ & $  35.9^{+  3.2}_{-  4.0}$ & $  3.8\pm 0.1$ & $ 13.4^{+ 1.9}_{-  1.6}$ & $ ~   7.6^{+  4.0}_{-  3.7}$ & $  43 $ & $ 45.7/  41$ \\  \noalign{\vskip 2pt}
32 & $  1.06^{+ 0.03}_{- 0.02}$ & $  10.6^{+  1.0}_{-  1.4}$ & $  3.7\pm 0.0$ & $ 12.9\pm 0.9$ & $ ~   9.0^{+  3.2}_{-  2.9}$ & $  48 $ & $ 71.7/  41$ \\  \noalign{\vskip 2pt}
\end{tabular}
\end{center}
{\small 
See Tab.~\ref{tab:x3fits} for an explanation of the fit
parameters.
The interstellar equivalent
column was fixed at $N_{\rm H}=7.2\times 10^{21}\,\rm cm^{-2}$. 
}
\end{table*}